\documentclass[twocolumn,prd,amsmath,superscriptaddress]{revtex4}

\usepackage{graphicx}
\usepackage{amsmath}
\usepackage{amssymb}
\usepackage{graphics}
\usepackage{hyperref}
\usepackage{ifthen}
\usepackage{bm}
\usepackage[usenames,dvipsnames]{color}

\newcommand{\DEs}{\Delta E_\mathrm{S}}
\newcommand{\DEt}{\Delta E_\mathrm{T}}
\newcommand{\dphithin}{\delta\phi_\mathrm{thin}}
\newcommand{\dVeff}{V_{\mathrm{eff},\phi}}
\newcommand{\DVloop}{\Delta V_\mathrm{1-loop}}
\newcommand{\Dxw}{\Delta x_\mathrm{w}}
\newcommand{\Dyw}{\Delta y_\mathrm{w}}
\newcommand{\Dzc}{\Delta z_\mathrm{C}}
\newcommand{\Dzfoil}{\Delta z_\mathrm{foil}}
\newcommand{\Dzl}{\Delta z_\mathrm{L}}
\newcommand{\Dzr}{\Delta z_\mathrm{R}}
\newcommand{\Dzst}{\Delta z_\textrm{S-T}}
\newcommand{\Dzw}{\Delta z_\mathrm{w}}
\newcommand{\Eh}{E_\mathrm{h}}
\newcommand{\Enh}{E_\mathrm{nh}}
\newcommand{\eotwash}{E\"ot-Wash}
\newcommand{\Etot}{E_\mathrm{tot}}
\newcommand{\Figa}{(a)}
\newcommand{\Figb}{(b)}
\newcommand{\Figbot}{(Bottom)}
\newcommand{\Figc}{(c)}
\newcommand{\Figd}{(d)}
\newcommand{\Figleft}{(Left)}
\newcommand{\Figright}{(Right)}
\newcommand{\Figtop}{(Top)}
\newcommand{\fsup}{f_\mathrm{sup}}
\newcommand{\meff}{m_\mathrm{eff}}
\newcommand{\Mpl}{M_\mathrm{Pl}}
\newcommand{\Nholes}{N_\mathrm{holes}}
\newcommand{\Nrows}{N_\mathrm{rows}}
\newcommand{\Nwedges}{N_\mathrm{wedges}}
\newcommand{\omegah}{\nu_\mathrm{h}}
\newcommand{\phiB}{\phi_\mathrm{B}}
\newcommand{\phiBl}{\phi_\mathrm{BL}}
\newcommand{\phiBr}{\phi_\mathrm{BR}}
\newcommand{\phic}{\phi_\mathrm{C}}
\newcommand{\phicone}{\phi_{\mathrm{C}1}}
\newcommand{\phictwo}{\phi_{\mathrm{C}2}}
\newcommand{\phig}{\phi_\mathrm{g}}
\newcommand{\phigone}{\phi_{\mathrm{g}1}}
\newcommand{\phigtwo}{\phi_{\mathrm{g}2}}
\newcommand{\phil}{\phi_\mathrm{L}}
\newcommand{\philone}{\phi_{\mathrm{L}1}}
\newcommand{\phir}{\phi_\mathrm{R}}
\newcommand{\phirtwo}{\phi_{\mathrm{R}2}}
\newcommand{\phis}{\phi_\mathrm{s}}
\newcommand{\phisv}{\phi_\mathrm{sv}}
\newcommand{\rhom}{\rho_\mathrm{m}}
\newcommand{\rhomc}{\rho_\mathrm{mC}}
\newcommand{\rhoml}{\rho_\mathrm{mL}}
\newcommand{\rhomr}{\rho_\mathrm{mR}}
\newcommand{\rhov}{\rho_\mathrm{v}}
\newcommand{\rhovone}{\rho_{\mathrm{v}1}}
\newcommand{\rhovtwo}{\rho_{\mathrm{v}2}}
\newcommand{\rsh}{r_\mathrm{Sh}}
\newcommand{\rth}{r_\mathrm{Th}}
\newcommand{\Veff}{V_\mathrm{eff}}
\newcommand{\zc}{z_\mathrm{c}}
\newcommand{\zg}{z_\mathrm{g}}
\newcommand{\Zgap}{{\mathcal Z}_\mathrm{gap}}
\newcommand{\zgone}{z_{\mathrm{g}1}}
\newcommand{\zgtwo}{z_{\mathrm{g}2}}

\begin{document}

\title{Dark energy fifth forces in torsion pendulum experiments}
\author{Amol Upadhye}
\affiliation{High Energy Physics Division, Argonne National Laboratory, 9700 S. Cass Ave., Argonne, IL 60439}%
\date{\today}

\begin{abstract}
The chameleon scalar field is a matter-coupled dark energy candidate whose nonlinear self-interaction partially screens its fifth force at laboratory scales.  Nevertheless, small-scale experiments such as the torsion pendulum can provide powerful constraints on chameleon models.  Here we develop a simple approximation for computing chameleon fifth forces in torsion pendulum experiments such as \eotwash.  We show that our approximation agrees well with published constraints on the quartic chameleon, and we use it to extend these constraints to a much wider range of models.  Finally, we forecast the constraints which will result from the next-generation \eotwash~experiment, and show that this experiment will exclude a wide range of quantum-stable models.
\end{abstract}

\maketitle

\section{Introduction}
\label{sec:introduction}


Evidence for an accelerating cosmic expansion is now solid~\cite{Komatsu_etal_2010,Larson_etal_2011,Suzuki_etal_2012,Sanchez_etal_2012}.  The simplest explanation for this acceleration, a ``cosmological constant'' vacuum energy density $\Lambda \Mpl^2$ which does not interact with Standard Model particles, remains consistent with the data; however, it must take an extremely small value $\sim 10^{-120}\Mpl^4$ which is difficult to explain without fine-tuning.  Alternative theories~\cite{Abbott_1985,Brown_Teitelboim_1987,Bousso_Polchinski_2000,Steinhardt_Turok_2006,Dvali_Hofmann_Khoury_2007,deRham_etal_2008,deRham_Hofmann_Khoury_Tolley_2008,Agarwal_etal_2009,Lehners_Steinhardt_2009,Lehners_Steinhardt_Turok_2009,Khoury_Steinhardt_2011} explain the smallness of this number dynamically, either through tunneling to a low-energy vacuum or through a slow reduction of the vacuum energy known as ``degravitation.''  Since the simplest of these models reduce at low energies to effective scalar field theories possibly coupled to known particles, it is interesting to consider such a scalar field ``dark energy'' independently of these more fundamental theories, and to ask how it may differ from a cosmological constant.  Generally speaking, such differences take two forms: a slow evolution of the background energy density~\cite{Peebles_Ratra_1988,Ratra_Peebles_1988}; and couplings to Standard Model particles, which we consider here.

Large fifth forces have not been observed, so coupled dark energy must have some mechanism to screen such couplings at laboratory and solar system scales.  Galileon fields invoke the nonlinear Vainshtein mechanism to reduce their effective couplings at high densities~\cite{Vainshtein_1972,Nicolis_Rattazzi_Trincherini_2008,Chow_Khoury_2009}.  Symmetron models decouple from matter through a symmetry restoration at high densities, while fifth forces exist in a symmetry-broken phase at low densities~\cite{Hinterbichler_Khoury_2010,Hinterbichler_Khoury_Levy_Matas_2011,Brax_etal_2012}.  The first screened scalars to be considered as dark energy candidates are chameleon models, which evade constraints by becoming massive in high-density environments~\cite{Khoury_Weltman_2004a,Khoury_Weltman_2004b,Brax_etal_2004}.  
The current article focuses on chameleon models.

Although these scalar fields are screened, such screening mechanisms are not perfect.  Laboratory experiments are powerful probes of residual fifth forces and new particles which could result from coupled dark energy~\cite{Adelberger_Heckel_Nelson_2003,Adelberger_etal_2009}.  Particles of a photon-coupled scalar could be produced through oscillation in a background magnetic field and detected using ``afterglow'' experiments~\cite{Chou_etal_2009,Brax_etal_2007b,Ahlers_etal_2008,Gies_Mota_Shaw_2008,Steffen_etal_2010,Upadhye_Steffen_Weltman_2010,Upadhye_Steffen_Chou_2012,Brax_Burrage_Davis_2012}.  Fifth forces may be probed directly through small-scale tests of gravity such as torsion pendulum experiments and Casimir force measurements~\cite{Adelberger_etal_2009,Fischbach_Talmadge_1999,Long_etal_2003,Kapner_etal_2007,Brax_etal_2007c,Weld_etal_2008,Kreuz_etal_2009,Brax_etal_2010b,Brax_Pignol_2011,Bassler_etal_2012}.

The goal of this article is to use torsion pendulum experiments such as \eotwash~\cite{Kapner_etal_2007} to constrain fifth forces from chameleon dark energy models~\cite{Gubser_Khoury_2004,Upadhye_Gubser_Khoury_2006,Mota_Shaw_2006,Mota_Shaw_2007,Adelberger_etal_2007,Brax_etal_2007c,Brax_etal_2010b,Upadhye_Hu_Khoury_2012}.  Previous work~\cite{Adelberger_etal_2007} used the numerical computations of~\cite{Upadhye_Gubser_Khoury_2006} to calculate the three-dimensional field configuration directly for the geometry of the \eotwash~experiment, a powerful probe of gravitation-strength fifth forces at submillimeter scales.  In this work we develop a simple, accurate estimate of the field profile for such experiments by approximating the matter distribution locally as one-dimensional and planar.  This one-dimensional plane-parallel (1Dpp) approximation allows us to compute the field on the surface of the source and test masses in a torsion pendulum experiment, from which the energy and torque can be found.  We show that the 1Dpp approximation agrees with the numerical calculations of~\cite{Upadhye_Gubser_Khoury_2006} and the data analysis of~\cite{Adelberger_etal_2007} for \eotwash, and we estimate the constraints on a much wider range of models.  

Recently it was shown that a subset of chameleon models is ``quantum-stable'' in the sense of having small one-loop corrections to the effective mass and bulk field value~\cite{Upadhye_Hu_Khoury_2012}.  For gravitation-strength couplings, quantum-stable models lie just beyond the bounds of \eotwash.  Using our 1Dpp approximation, we forecast constraints from the next-generation \eotwash~experiment and show that it rules out a large range of such models.

The paper proceeds as follows.  Section~\ref{sec:chameleon_fields} introduces the chameleon model and its fifth force screening mechanism.  In Sec.~\ref{sec:field_profile_in_planar_systems} we study in detail the one-dimensional planar problem, which is exactly solvable for power law chameleon potentials.  Solutions of this one-dimensional problem are used to approximate the expected torsion pendulum signal in Sec.~\ref{sec:torsion_pendulum_experiments}, and Sec.~\ref{sec:conclusion} concludes.

\section{Chameleon fields}
\label{sec:chameleon_fields}

\subsection{Equation of motion}
\label{subsec:equation_of_motion}

The chameleon field $\phi$ is a canonically normalized scalar field with a nonlinear self-interaction and a matter coupling~\cite{Khoury_Weltman_2004a,Khoury_Weltman_2004b,Brax_etal_2004}.  A simple matter interaction results from the conformal coupling of the chameleon field to the metric, as given by the action
\begin{equation}
S
=
\int d^4x 
\left( 
-\frac{(\partial \phi)^2 }{2}
- 
V(\phi)
+
{\mathcal L}_\mathrm{m}(\psi_i,e^\frac{2\beta\phi}{\Mpl} g_{\mu\nu})
\right)
\label{e:S}
\end{equation}
in the flat-spacetime case appropriate to laboratory tests.  Here $V(\phi)$ is the chameleon potential, and matter is represented as Fermion fields $\psi_i$ with Lagrangian density ${\mathcal L}_m$.  Conformal coupling of the chameleon results in a universal coupling constant $\beta>0$ to all Fermionic matter, a feature which is stable with respect to quantum corrections~\cite{Fujii_1997,Hui_Nicolis_2010}.  In a background matter density $\rho(\vec x)$ the chameleon equation of motion is 
\begin{eqnarray}
\partial_\mu \partial^\mu \phi
&=&
\frac{\partial \Veff}{\partial \phi}
\\
\Veff(\phi,\vec x)
\label{e:eom}
&=&
V(\phi)
+
\frac{\beta}{\Mpl} \rho(\vec x) \phi,
\label{e:Veff}
\end{eqnarray}
where $\Veff$ is the effective potential.  We have neglected terms of higher order in $\beta \phi / \Mpl$ since this quantity will be small in all models of interest.   

In this work we will primarily be concerned with the static case, in which the equation of motion reduces to
\begin{equation}
\nabla^2 \phi
=
V'(\phi)
+
\beta \rho / \Mpl.
\label{e:eom_static}
\end{equation}
Deep inside a bulk medium of constant density $\rho_0$ even the spatial derivatives vanish.  The bulk field $\phiB(\rho_0)$ then satisfies $V'(\phiB(\rho_0)) + \beta\rho_0/\Mpl = 0$.  The effective mass associated with small fluctuations about a field $\phi$ is given by $\meff(\phi)^2 = V''(\phi)$.

The chameleon potential $V(\phi)$ must be chosen to fit the cosmological data and to screen fifth forces locally.  Cosmology requires that $V>0$ vary sufficiently slowly with time, and we will see that the chameleon effect requires $V'<0$ and $V''>0$.  Since the cosmic acceleration is sourced by constant or slowly-varying parts of $V$ while laboratory experiments are sensitive only to derivatives of $V$, we choose a constant-plus-power-law potential which splits these two regimes:
\begin{equation}
V(\phi)
=
M_\Lambda^4\left( 1 + \gamma \left|\frac{\phi}{M_\Lambda}\right|^n \right).
\label{e:V_chamDE}
\end{equation}
Here $M_\Lambda = 2.4 \times 10^{-3}$~eV is the dark energy scale, so that the constant term $M_\Lambda^4$ drives the cosmic acceleration.  The second term, in which $\gamma>0$ and $n$ are dimensionless numbers, can be probed in laboratory experiments.  For $n=4$ it is conventional to define $\lambda \equiv 4! \gamma$.  The bulk field and mass are given by
\begin{eqnarray}
\phiB(\rho)
&=&
\sigma_n M_\Lambda \left(\frac{\beta\rho}{|n|\gamma M_\Lambda^3 \Mpl}\right)^\frac{1}{n-1}
\label{e:phi_bulk}
\\
\meff(\rho)
&=&
M_\Lambda |n-1|^\frac{1}{2} (|n|\gamma)^\frac{1}{2n-2}
\left(\frac{\beta\rho}{M_\Lambda^3 \Mpl}\right)^\frac{n-2}{2n-2}
\label{e:meff_bulk}
\end{eqnarray}
where $\sigma_n = \mathrm{sign}(1-n)$.  

This potential is the large-field limit of the exponential potential $V = M_\Lambda^4 \exp(\gamma \phi^n/M_\Lambda^n)$ of ~\cite{Brax_etal_2004}.  Henceforth we work with (\ref{e:V_chamDE}) whenever specific examples or constraints are presented.  Furthermore, the mass does not increase with density when $1 < n < 2$, and we will see that (\ref{e:V_chamDE}) is constrained by cosmology when $-1/2 \lesssim n < 1$.  Thus we only consider models with $n \lesssim -1/2$ or $n>2$.  Note that, due to our sign convention $\beta > 0$, the field $\phi$ will be negative for $n>2$ and positive for $n<0$. For all such $n$, $\phiB(\rho)$ decreases as $\rho$ increases.

\subsection{Chameleon and thin-shell effects}
\label{subsec:chameleon_and_thin-shell_effects}

Chameleon phenomenology is characterized by the presence of two regimes: a linear, or ``unscreened,'' regime; and a nonlinear, ``screened'' regime.  In the linear regime, the potential derivative term on the right hand side of (\ref{e:eom_static}) is negligible, so the equation of motion is approximately linear in $\phi$.  Furthermore, the source term remaining on the right hand side is proportional to $\rho$; thus (\ref{e:eom_static}) in the linear regime is similar to the Poisson equation $\nabla^2 \Psi = \rho / (2\Mpl^2)$ for the gravitational potential $\Psi$.  Since gradients of $\phi$ and $\Psi$ vanish far from an object, $\phi$ is equal to $2\beta \Mpl \Psi$ up to an additive constant,
\begin{equation}
\Delta \phi^\mathrm{(lin)} 
=
2\beta \Mpl \, \Delta \Psi
\label{e:Dphi_lin}
\end{equation}
where the $\Delta$ denotes a difference between two spatial positions.  The linear regime applies, for example, to a dense object of sufficiently small volume in a sufficiently low-density bulk.  

Suppose that the volume of such an object is increased at constant density.  For a characteristic size $r$ the gravitational potential $\Psi \propto \rho r^2$, and $\phi$ will change with $\Psi$ throughout the linear regime. As $\phi$ changes from its minimum, $V'(\phi)$ will become large and negative, partially cancelling the source density on the right side of  (\ref{e:eom_static}).  

This cancellation, known as ``screening'' of the source, is characteristic of the nonlinear regime of chameleon models.  In the nonlinear limit this screening becomes complete and the field asymptotically approaches its bulk value $\phiB(\rho)$.  Since the gravitational potential continues to grow linearly, in the nonlinear regime the change in the field value will be much smaller than the linear approximation~(\ref{e:Dphi_lin}),
\begin{equation}
\left|\Delta\phi^\mathrm{(nl)}\right|
\ll
2\beta \Mpl \, \left|\Delta \Psi\right|.
\label{e:Dphi_nl}
\end{equation}

Suppose that the chameleon field at the center of an object in the nonlinear regime is $\phi_0$, and the field far away is $\phi_\infty$.  The gravitational potential at the center is $\Psi_0$; far away, $\Psi$ is defined to be zero.  For chameleon models with negative $n$, $\phi_0 \approx 0$, so (\ref{e:Dphi_nl}) becomes $|\phi_\infty| \ll 2\beta \Mpl |\Psi_0|$.  For $n>2$, the opposite is true; $\phi_0 \gg \phi_\infty$, so (\ref{e:Dphi_nl}) implies $|\phi_0| \ll 2\beta \Mpl |\Psi_0|$.  The gravitational potential of a disk of radius $r_\mathrm{disk}$ and thickness $z_\mathrm{disk}$ is approximately $\Psi \sim \rho r_\mathrm{disk} z_\mathrm{disk}$; Ref.~\cite{Brax_etal_2007c} finds $\Psi = \rho r_\mathrm{disk} z_\mathrm{disk} / (8\Mpl^2)$ to be a good approximation.

The chameleon fifth force in the nonlinear regime is suppressed by two effects known as the \emph{chameleon} and \emph{thin-shell} effects.  The chameleon effect is the rapid growth of the effective mass $\meff$ with the size and density of a source object, which effectively converts a long-range force into a short-range one.  For example, the Compton wavelength of a $\gamma=1$, $n=-1$, $\beta=1$ model increases from $\sim 100$~pc at cosmological densities to $\sim 0.1$~mm at laboratory densities.  The thin-shell effect is due to the screening of the interior of a source mass.  If the chameleon field is near its bulk value inside an object, and changes only in a thin shell of matter at the outer edge of that object, then a test mass outside that object will ``see'' only the fifth force due to that thin shell of matter.  Due to the thin-shell effect, a chameleon with a gravitation-strength coupling $\beta \sim 1$ can easily be consistent with solar system fifth force constraints.  Because of the chameleon and thin-shell effects, a model with effective mass $\meff$ at a given density is best probed using objects of size $\sim \meff^{-1}$ separated by a distance $\sim \meff^{-1}$.

\begin{figure}[t]
\begin{center}
\includegraphics[angle=270,width=2.8in]{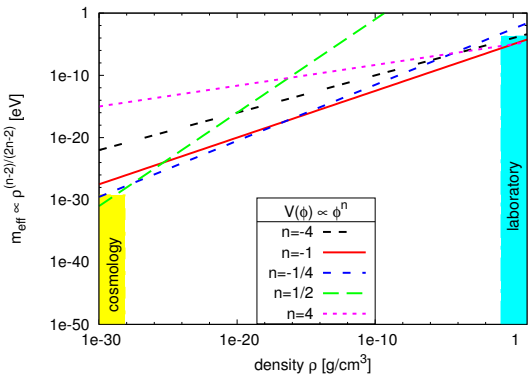}
\caption{Mass scalings of various chameleon models.  Laboratory constraints are most suitable for probing models with $n \lesssim -1/2$ and $n > 2$, while cosmological fifth force constraints probe the models $-1/2 \lesssim n < 1$ in which $\meff$ grows most rapidly with density.\label{f:which_scale}}
\end{center}
\end{figure}

Using the mass scaling~(\ref{e:meff_bulk}), $\meff \propto \rho^{(n-2)/(2n-2)}$, we can estimate whether a given chameleon  model can best be constrained by laboratory or cosmological experiments.  Small-scale tests of gravity can exclude unscreened gravitation-strength fifth forces on length scales $\gtrsim 1$~mm at densities $\rho \sim 1$~g/cm$^3$, while cosmological probes exclude such forces on megaparsec length scales at the cosmic background density $\sim 10^{-30}$~g/cm$^3$.  These approximate excluded regions are shaded in Figure~\ref{f:which_scale}.  Suppose we have a model which can barely be probed in the laboratory, $\meff \sim (1\textrm{ mm})^{-1}$, such as the models with $n=4$, $n=-1$, and $n=-4$ shown in the figure.  This model will be too massive to probe cosmologically if the mass at cosmological densities is greater than $(1\textrm{ Mpc})^{-1}$.  This condition is satisfied if $\frac{n-2}{n-1} < \frac{1}{15} \log_{10}\frac{1\textrm{ Mpc}}{1\textrm{ mm}}$, implying either $n \lesssim -1/2$ or $n>1$.  The remaining models have rapid mass scalings and are best probed cosmologically.

\subsection{Quantum stability condition}
\label{subsec:quantum_stability_condition}

Reference~\cite{Upadhye_Hu_Khoury_2012} derived a condition for the quantum stability of a chameleon potential, that is, the condition that quantum corrections to the potential be subdominant over the range of field values probed by a fifth force experiment.  Large masses, which help a chameleon model to evade fifth force constraints, also lead to large quantum corrections.  For a general class of potentials, Ref.~\cite{Upadhye_Hu_Khoury_2012} found the quantum stability condition $\meff(\rho) < 0.0073 (\rho/10\textrm{ g/cm}^3)^{1/3}$~eV.  Quantum-stable models are an interesting subset of all chameleon models, and we will see that the next-generation \eotwash~experiment can exclude a range of quantum-stable chameleons.

For a potential such as~(\ref{e:V_chamDE}), the quantum stability condition is the requirement that the one-loop Coleman-Weinberg corrections to the slope of the potential and to the chameleon effective mass  (that is to $V'$ and $V''$) be smaller in magnitude than their tree-level counterparts.  The one-loop correction to the potential is
\begin{equation}
\DVloop
=
\frac{V''(\phi)^2}{64\pi^2} \log\left(\frac{V''(\phi)}{\mu_0^2}\right)
\label{e:DVloop}
\end{equation}
where  primes denote derivatives of the potential (\ref{e:V_chamDE}) with respect to $\phi$, and $\mu_0$ is a mass scale which characterizes the chameleon field in the experiment.  Corrections to $V'$ and $V''$ are the first and second derivatives of (\ref{e:DVloop}), respectively, so the quantum stability conditions are 
\begin{equation}
\left|\frac{\DVloop'}{V'}\right|, \, \left|\frac{\DVloop''}{V''}\right| < 1.
\label{e:quantum_stability}  
\end{equation}
Although we can choose $\mu_0$ to make $\DVloop$ zero at any given field value, a fifth force experiment will probe a range of field values.  $\phi$ and $\meff$ can vary by an order of magnitude or more in a typical experiment, so quantum corrections will not be zero everywhere.  When we consider the quantum stability of a model in a specific experimental setup in Section~\ref{sec:torsion_pendulum_experiments}, we will choose $\mu_0$ from among the chameleon masses in the experiment so as to minimize quantum corrections.

As an estimate of quantum corrections, we may set the log term in (\ref{e:DVloop}) to unity and evaluate~(\ref{e:quantum_stability}) at the bulk field $\phiB(\rho_0)$ corresponding to some density $\rho_0$.
For the potential (\ref{e:V_chamDE}) the stability conditions are 
\begin{eqnarray}
\gamma^\frac{3}{n-1}
\left(\frac{\beta\rho_0}{|n|M_\Lambda^3 \Mpl}\right)^\frac{n-4}{n-1}
&<&
\frac{32 \pi^2}
     {|n(n-1)^2(n-2)|}
\label{e:quantum_stability_1}
\\
\gamma^\frac{3}{n-1}
\left(\frac{\beta\rho_0}{|n|M_\Lambda^3 \Mpl}\right)^\frac{n-4}{n-1}
&<&
\frac{32 \pi^2}
     {|n(n-1)(n-2)(2n-5)|}\quad\,\,\,
\label{e:quantum_stability_2}
\end{eqnarray}
Note that quantum stability imposes an upper bound on the self-coupling $\gamma$ for $n>2$ and a lower bound for $n<0$.  In the case $n=4$ the density-dependent term disappears, and (\ref{e:quantum_stability_1}-\ref{e:quantum_stability_2}) imply $4!\cdot\gamma = \lambda < 32\pi^2/3 \approx 105$.  Casimir force constraints can rule out quantum-stable $n=4$ chameleons with strong matter couplings $\beta \gtrsim 10^6$~\cite{Brax_etal_2007c}; however, quantum-stable models with gravitation-strength couplings $\beta \sim 1$ remain allowed.

\section{Field profile in planar systems}
\label{sec:field_profile_in_planar_systems}

\subsection{Planar slab in vacuum}
\label{subsec:planar_slab_in_vacuum}

The field profile in the vacuum outside an infinitely thick planar slab can be found exactly~\cite{Upadhye_Gubser_Khoury_2006,Brax_etal_2007c}.  Let $\rho(z) = \rho_0 \Theta(-z)$ where $\Theta$ is the step function.  Thus $\rho$ is positive for $z<0$ and zero for $z>0$; the face of the slab is the $xy$ plane, and its normal is $\hat z$.  The static equation of motion (\ref{e:eom_static}) in the vacuum, $d^2\phi/dz^2 = dV/d\phi$, is solved for the potential (\ref{e:V_chamDE}) by 
\begin{equation}
\phi(z) 
=
\phisv \left(1 + \sqrt{\frac{1}{2} (n-2)^2 \gamma M_\Lambda^{4-n}} \phisv^\frac{n-2}{2} z\right)^{-\frac{2}{n-2}}
\label{e:phi_vacuum_exterior}
\end{equation}
where $\phisv$ is the field value on the surface $z=0$ of the slab in vacuum.  Using $\frac{d^2\phi}{dz^2} = \frac{1}{2}\frac{d}{d\phi}\frac{d\phi}{dz}$ to integrate the equation of motion, we find
\begin{equation}
\frac{1}{2}\left.\left(\frac{d\phi}{dz}\right)^2\right|_{\phi_i}^{\phisv}
=
\Veff(\phisv,\rho) - \Veff(\phi_i,\rho).
\label{e:eom_1D_integrated}
\end{equation}
Choosing $\phi_i = \phiB(\rho_0)$, corresponding to $z\rightarrow -\infty$ and $\rho = \rho_0$, yields one equation relating $\phisv$ to $d\phi(0)/dz$; choosing $\phi_i = \phiB(0)$, corresponding to $z \rightarrow \infty$ and $\rho = 0$, yields another.   Combining the two, and noting that $d\phi/dz\rightarrow 0$ as $z \rightarrow \pm \infty$, gives the simple result
\begin{equation}
\phisv 
=
\left(1 - \frac{1}{n}\right) \phiB(\rho_0).
\end{equation}

An exact, closed-form solution is not available inside the thick slab. However, we can linearize the equation of motion around $\phiB(\rho_0)$ and require $\phi(z)$ to be continuous at $z=0$:
\begin{equation}
\phi_\mathrm{thick}(z) 
\approx 
\phiB(\rho_0) + [\phisv-\phiB(\rho_0)] e^{\meff(\rho_0)z}.
\label{e:thick-slab_linearization}
\end{equation}

The case of a thin slab is slightly more complicated.  Suppose that the slab is centered at $z=\zc$ with a half-thickness of $\delta z$.  Guess a value $\phic = \phi(\zc)$.  The equation of motion linearized about $\phic$, and its solution $\dphithin(z) = \phi(z) - \phic$ inside the slab, are
\begin{eqnarray}
\frac{d^2 \delta\phi}{dz^2}
&\approx&
\dVeff(\phic,\rho_0) + \meff(\phic)^2\delta\phi
\\
\Rightarrow
\dphithin(z)
&\approx&
\frac{\dVeff(\phic,\rho_0)}{\meff(\phic)^2}
[\cosh(\meff(\phic)z)-1]. \qquad
\label{e:thin-slab_linearization}
\end{eqnarray}
Replacing $\phisv$ by $\phic + \delta\phi_\mathrm{thin}(\zc + \delta z)$ in (\ref{e:phi_vacuum_exterior}), we can find the exterior solution corresponding to this guess $\phic$.  When the correct value of $\phic$ is chosen, $d\phi/dz$ as well as $\phi$ will be continuous at $\zc + \delta z$.  However, if we consider thicker and thicker slabs, we cannot be sure that such a solution will exist.

\subsection{Planar gap}
\label{subsec:planar_gap}

Consider a planar gap with $\rho=\rhov$ bounded on the left, $z\leq 0$, by a thick slab of density $\rhoml$, and on the right, $z \geq \Delta z$, by a thick slab of density $\rhomr$.  That is, $\rho(z) = \rhoml \Theta(-z) + \rhov \Theta(z)\Theta(\Delta z - z) + \rhomr \Theta(z-\Delta z)$.  There are four unknowns: the surface field values $\phil$ and $\phir$ at $z=0$ and $\Delta z$, respectively; the maximum field value $\phig$ inside the gap; and $\zg$, the point at which $\phi = \phig$.  

Equation~\ref{e:eom_1D_integrated} can be applied to any interval $[\phi(z_i),\phi(z_j)]$ over which $\rho$ is constant.  The intervals $[\phiB(\rhoml),\phil]$ and $[\phil,\phig]$ give, respectively,
\begin{eqnarray}
\frac{1}{2}\left.\left(\frac{d\phi}{dz}\right)^2\right|_{\phil}
&=&
V(\phil) - V(\phiB(\rhoml)) 
\nonumber\\
&\quad&
+ 
\frac{\beta \rhoml (\phil-\phiB(\rhoml))}{\Mpl} 
\\
-\frac{1}{2}\left.\left(\frac{d\phi}{dz}\right)^2\right|_{\phil}
&=&
V(\phig) - V(\phil)
+
\frac{\beta\rhov(\phig-\phil)}{\Mpl}.  \qquad
\end{eqnarray}
Adding the two yields a relation between $\phil$ and $\phig$.  A similar procedure can be applied to the plane on the right.  Thus we have
\begin{eqnarray}
\phil
&=&
\frac{\rhoml \phiB(\rhoml) - \rhov \phig}{\rhoml-\rhov}
+
\frac{V(\phiB(\rhoml)) - V(\phig)}{\beta \Mpl^{-1} (\rhoml-\rhov)}
\\
\phir
&=&
\frac{\rhomr \phiB(\rhomr) - \rhov \phig}{\rhomr-\rhov}
+
\frac{V(\phiB(\rhomr)) - V(\phig)}{\beta \Mpl^{-1} (\rhomr-\rhov)}. \qquad
\end{eqnarray}

Next, we apply (\ref{e:eom_1D_integrated}) to $[\phi(z),\phig]$ for some arbitrary $z$ between $0$ and $\zg$ in order to find $d\phi/dz$ inside the gap:
\begin{equation}
\frac{d\phi}{dz}
=
\sqrt{2}
\sqrt{V(\phi) - V(\phig) + \frac{\beta\rhov}{\Mpl}(\phi-\phig)}.
\label{e:1Dpp_dphidz}
\end{equation}
We can integrate with respect to $\phi$ between $\phil$ and $\phig$ to determine $\zg$ in terms of $\phig$:
\begin{eqnarray}
\zg
&=&
\int_{\phil}^{\phig} 
\frac{d\phi / \sqrt{2 \gamma M_\Lambda^{4-n}}}
     {\sqrt{|\phi|^n - |\phig|^n - |n||\phiB(\rhov)|^{n-1}(\phig-\phi)}}
\nonumber\\
&=&
\sum_{i=0}^\infty \sum_{j=0}^i 
\frac{(-1)^j \sigma_n^i \Gamma(\frac{1}{2})(|n| |\phiB(\rhov)|^{n-1})^i \phig^{i-j}}
     { \Gamma(\frac{1}{2}-i) j! (i-j)!\sqrt{2 \gamma M_\Lambda^{4-n}}}
\nonumber\\
&\quad&
\times
\int_{|\phil|}^{|\phig|}
\frac{|\phi|^j \, d|\phi|}
     {(|\phi|^n - |\phig|^n)^{i+1/2}}
\nonumber\\
&=&
\frac{\sqrt{\frac{n-1}{2n}}}{\meff(\phig)}
\sum_{i=0}^\infty \sum_{j=0}^i
\frac{(-1)^j \sigma_n^i \Gamma(\frac{1}{2})|n|^i}
     {\Gamma(\frac{1}{2}-i) j! (i-j)!}
\left|\frac{\phiB(\rhov)}{\phig}\right|^{i(n-1)}
\nonumber\\
&\quad&
\times
{\mathcal B}_{1-\left|\frac{\phig}{\phil}\right|^n}
\left(\frac{1}{2}-i, \frac{1}{2} + i - \frac{1+j}{n}\right)
\nonumber\\
&\equiv&
\Zgap(\phig,\phil,\rhov)
\label{e:1Dpp_1gap_Dzleft}
\end{eqnarray}
where ${\mathcal B}_x(a,b) = \int_0^x t^{a-1} (1-t)^{b-1}dt$ is the incomplete Beta function, and we have defined the shorthand $\Zgap$ for this expression as a function of the gap field $\phig$, surface field $\phil$, and gap density $\rhov$.
Repeating this procedure for the right side of the gap, $\zg < z < \Delta z$, 
\begin{equation}
\Delta z - \zg
=
\Zgap(\phig,\phir,\rhov),
\label{e:1Dpp_1gap_Dzright}
\end{equation}
which is similar to (\ref{e:1Dpp_1gap_Dzleft}) but with $\phil$ replaced by 
$\phir$ in the incomplete Beta function.

Thus we have $\phil$, $\phir$, and $\zg$ in terms of $\phig$, while the sum of (\ref{e:1Dpp_1gap_Dzleft}) and (\ref{e:1Dpp_1gap_Dzright}) implicitly defines $\phig$ as a function of the known gap size $\Delta z$.  By guessing $\phig$, comparing the resulting $\Delta z$ to the known value, and iteratively refining our guess, we can find $\phig$.  Once $\phig$ is known, we can integrate (\ref{e:1Dpp_dphidz}) to find $\phi(z)$ within the gap.  For example, given any $\phi_0=\phi(z_0)$ between $\phil$ and $\phig$, we obtain for $\zg - z_0$ a formula similar to (\ref{e:1Dpp_1gap_Dzleft}) with $\phil$ replaced by $\phi_0$.

The series sums in (\ref{e:1Dpp_1gap_Dzleft},~\ref{e:1Dpp_1gap_Dzright}) will converge quickly unless $\phig$ is close to $\phiB(\rhov)$.  In that case, the fifth force on each slab will be small anyway; the gap is large enough that the opposite slab does not pull $\phig$ very far from its bulk vacuum value.   When fifth forces are large, even the $i=0$ term alone is a reasonable approximation: $\sqrt{\frac{2n}{n-1}}\meff(\phig)\Delta z \approx {\mathcal B}_{1-(\phig/\phil)^n}(1/2,1/2-1/n) + {\mathcal B}_{1-(\phig/\phir)^n}(1/2,1/2-1/n)$.  Henceforth, in our numerical calculations, we truncate $\Zgap$ after the $i=5$ terms.  

To summarize, we have shown how to compute the surface field $\phir(\Delta z)$ as a function of gap size $\Delta z$ in a planar system.  This result will be essential to our approximation for torsion pendulum experiments in Sec.~\ref{sec:torsion_pendulum_experiments}.

\subsection{Thin slab in planar gap}
\label{subsec:thin_slab_in_planar_gap}

Before proceeding to experiments we study one final planar configuration, the thin slab inside a planar gap.  This will allow us to
calculate the chameleon screening caused by the electrostatic shielding foil between source and test masses in short-range fifth force experiments.

\begin{figure}[t]
\begin{center}
\includegraphics[angle=270,width=3.4in]{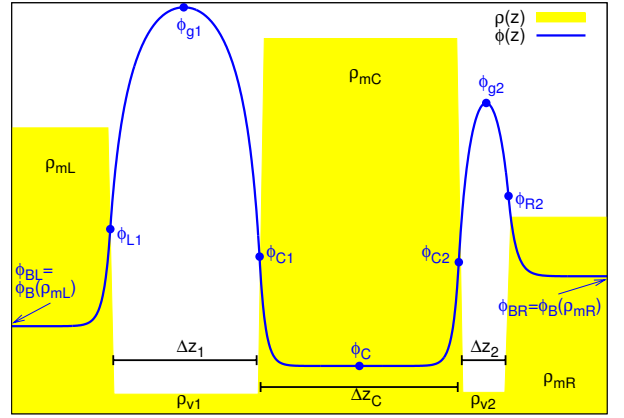}
\caption{Density (shaded yellow region) and field profile (solid blue line) for a thin planar slab inside a gap between two thick planar slabs.  The horizontal axis shows the $z$ coordinate, the distance normal to the planes, while the vertical axis shows $\rho$ and $\phi$ in arbitrary units.  Distances, densities, and field values are labeled.  \label{f:planar_gap_3slabs}}
\end{center}
\end{figure}

Figure~\ref{f:planar_gap_3slabs} shows the geometry considered here.  A central slab of width $\Dzc$ and density $\rhomc$ sits between two thick slabs, one on the left at a distance $\Dzl$ with a density $\rhoml$, and another on the right at a distance $\Dzr$ with a density $\rhomr$.  Gap 1, between the left and central slabs, has a ``vacuum'' with density $\rhovone$, while gap 2, between the central and right slabs, has a densty $\rhovtwo$; we assume $\rhovone, \rhovtwo \ll \rhoml, \rhomc, \rhomr$.  Field values deep inside the left and right slabs are $\phiBl = \phiB(\rhoml)$ and $\phiBr = \phiB(\rhomr)$, respectively.  Without loss of generality, let $z=0$ be the face of the left slab.  Given these values, we wish to find: $\philone = \phi(0)$; $\phigone$, the maximum inside gap 1; $\phicone = \phi(\Dzl)$; $\phic$, the minimum inside the central slab; $\phictwo = \phi(\Dzl+\Dzc)$; $\phigtwo$, the maximum inside gap 2; $\phirtwo = \phi(\Dzl+\Dzc+\Dzr)$; and the positions $\zgone$, $\zc$, and $\zgtwo$ at which the local extrema $\phigone$, $\phic$, and $\phigtwo$, respectively, are attained.

For these ten unknowns we have ten equations.  Two are obtained by applying (\ref{e:eom_1D_integrated}) to itervals $[\philone,\phigone]$ and $[\phigone,\phicone]$; two more by evaluating the thin-slab linearization (\ref{e:thin-slab_linearization}) at $z = \Dzl$ and $\Dzl+\Dzc$; and two more by applying (\ref{e:eom_1D_integrated}) to itervals $[\phictwo,\phigtwo]$ and $[\phigtwo,\phirtwo]$.  The final four are found by applying (\ref{e:1Dpp_1gap_Dzleft}, \ref{e:1Dpp_1gap_Dzright}) to the intervals $[0,\zgone]$, $[\zgone,\Dzl]$, $[\Dzl+\Dzc,\Dzl+\Dzc+\zgtwo]$, and $[\Dzl+\Dzc+\zgtwo,\Dzl+\Dzc+\Dzr]$, that is, to the left and right sides of gaps 1 and 2.
\begin{eqnarray}
0
&=&
\left.\Veff(\phi,\rhoml)\right|_{\phiBl}^{\philone}
+
\left.\Veff(\phi,\rhovone)\right|_{\philone}^{\phigone}
\label{e:1Dpp_3slabs_I}
\\
0
&=&
\left.\Veff(\phi,\rhovone)\right|_{\phigone}^{\phicone}
+
\left.\Veff(\phi,\rhomc)\right|_{\phicone}^{\phic}
\label{e:1Dpp_3slabs_II}
\end{eqnarray}
\begin{eqnarray}
\phicone
&=&
\phic
+
\dphithin(\Dzl)
\label{e:1Dpp_3slabs_III}
\\
\phictwo
&=&
\phic
+
\dphithin(\Dzl+\Dzc)
\label{e:1Dpp_3slabs_IV}
\end{eqnarray}
\begin{eqnarray}
0
&=&
\left.\Veff(\phi,\rhomc)\right|_{\phic}^{\phictwo}
+
\left.\Veff(\phi,\rhovtwo)\right|_{\phictwo}^{\phigtwo}
\label{e:1Dpp_3slabs_V}
\\
0
&=&
\left.\Veff(\phi,\rhovtwo)\right|_{\phigtwo}^{\phirtwo}
+
\left.\Veff(\phi,\rhomr)\right|_{\phirtwo}^{\phiBr}
\label{e:1Dpp_3slabs_VI}
\end{eqnarray}
\begin{eqnarray}
\zgone
&=&
\Zgap(\phigone,\philone,\rhovone)
\label{e:1Dpp_3slabs_VII}
\\
\Dzl
&=&
\Zgap(\phigone,\phicone,\rhovone)
+
\zgone
\label{e:1Dpp_3slabs_VIII}
\\
\zgtwo
&=&
\Zgap(\phigtwo,\phictwo,\rhovtwo)
+
\Dzl+\Dzc
\label{e:1Dpp_3slabs_IX}
\\
\Dzr
&=&
\Zgap(\phigtwo,\phirtwo,\rhovtwo)
+
\zgtwo
-\Dzl-\Dzc
\label{e:1Dpp_3slabs_X}
\end{eqnarray}

Note that if $\phigone$, $\phigtwo$, and $\phic$ are specified, then (\ref{e:1Dpp_3slabs_I}-\ref{e:1Dpp_3slabs_X}) immediately give the other seven unknowns.  We also obtain gap sizes $\Dzl'$ and $\Dzr'$ and slab thickness $\Dzc'$; however, these will not necessarily match the givens $\Dzl$, $\Dzr$, and $\Dzc$.  In order to find the correct $\phigone$, $\phigtwo$, and $\phic$, we minimize $(\Dzl'-\Dzl)^2+(\Dzr'-\Dzr)^2+(\Dzc'-\Dzc)^2$ numerically.

In the symmetric case, $\rhoml=\rhomr$, $\rhovone=\rhovtwo$, and $\Dzl=\Dzr$, the problem simplifies considerably.  Matching the field derivative at the surface of the central slab gives $(\dphithin'(\Dzc/2))^2/2 = \Veff(\phic+\dphithin(\Dzc/2),\rhovone) - \Veff(\phigone,\rhovone)$, where the prime ($'$) denotes $d/dz$.  Thus $\phigone$ determines $\phic$.  We need only solve numerically for the value of $\phigone$ which gives the right gap size $\Dzl$ using (\ref{e:1Dpp_3slabs_VII},~\ref{e:1Dpp_3slabs_VIII}).

\subsection{Force suppression due to shielding foil}
\label{subsec:force_suppression_due_to_shieldng_foil}

Forces at short ranges between small source and test masses in a fifth force experiment will typically be dominated by electrostatic effects.  Even electrostatic forces between stray charges can swamp gravitation-strength forces.  Thus most such experiments stretch a thin, grounded conducting foil between the source and test masses to shield the test mass from these electrostatic forces.  Such a foil can screen the chameleon fifth force as well, so we study it here.  

First, consider a system with two thick planar slabs.  Let the slab on the left be the source mass and the one on the right the test mass.  A change in the position of the source will change the surface value of the field on the test mass by some amount $\Delta \phi(\textrm{no-foil})$.  The change in the fifth force on the test mass, the experimental signal, is proportional to $\Delta \phi(\textrm{no-foil})$.

\begin{figure}[t]
\begin{center}
\includegraphics[angle=270,width=3.3in]{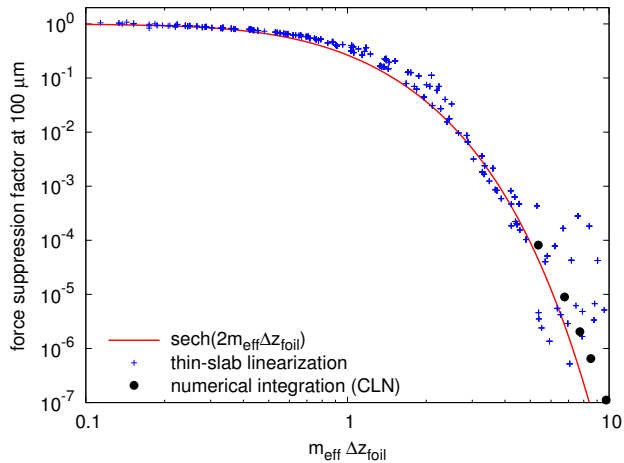}
\caption{Factor $\fsup$ by which the chameleon fifth force is suppressed by a shielding foil.  For a large range of chameleon parameters $\gamma$, $n$, and $\beta$, the suppression factor, shown by points on the plot, is a function of the foil thickness $\Dzfoil$ in Compton wavelengths.  $\fsup$ is well-approximated by the function $\textrm{sech}(2\meff z_\mathrm{foil})$ (solid line).   The thin-slab calculation of $\fsup$, from (\ref{e:1Dpp_3slabs_I}-\ref{e:1Dpp_3slabs_X}), is shown by blue ``+''-shaped points; the scatter at large $\meff \Dzfoil$ is due to numerical error, since $\Delta \phi(\mathrm{foil})$ is the difference of two nearly equal numbers.  Solid black circles show $\fsup$ computed directly from arbitrary-precision numerical integration of the equations of motion.  
\label{f:force_suppression_factor}}
\end{center}
\end{figure}

Now suppose that another slab, corresponding to the shielding foil, is placed between the source and test masses.  In the limit that this central slab is thick, it will completely screen chameleon fifth forces.  The field at its center will be close to its bulk value, and the field on the side facing the test mass will be very weakly dependent on the field at the opposite face.  In the case of a thin slab, however, this screening will not be total.  The results of Sec.~\ref{subsec:thin_slab_in_planar_gap} provide an excellent approximation to the surface field in the presence of a shielding foil. The change in source mass position will result in a change $\Delta \phi(\mathrm{foil})$ in the field on the surface of the test mass.  

Figure~\ref{f:force_suppression_factor} shows the suppression factor $\fsup \equiv \Delta \phi(\mathrm{foil}) / \Delta \phi(\textrm{no-foil})$ for a $\Dzc = 10$~$\mu$m foil at the center of a gap with $\Dzl+\Dzr+\Dzc = 100$~$\mu$m.  The density of each slab is $10$~g/cm$^3$ and the density in the gaps is $10^{-12}$g/cm$^3$, corresponding to air at room temperature and a pressure of $\lesssim 10^{-6}$~torr.  $\Delta \phi(\mathrm{foil})$ is found by varying $\Dzl$ by $10$~$\mu$m in either direction and using (\ref{e:1Dpp_3slabs_I}-\ref{e:1Dpp_3slabs_X}) to find the change in $\phirtwo$.  $\fsup$ is approximately equal to $\textrm{sech}(2 \meff \Dzc)$, where $\meff$ is evaluated at $\phiB(\rhomc)$.

At large $\meff \Dzfoil$, $\fsup$ is the difference between two nearly-equal numbers.  The resulting numerical error is responsible for the scatter in the ``+''-shaped points at $\meff \Dzfoil \gtrsim 5$ in Fig.~\ref{f:force_suppression_factor}.  In order to verify our approximation for $\fsup$ in this regime, we integrated the equation of motion numerically using the CLN arbitrary-precision arithmetic package~\cite{CLN}.  The resulting $\fsup$ values, shown as filled circles in the figure, agree well with $\fsup \approx \textrm{sech}(2\meff\Dzfoil)$.

\section{Torsion pendulum experiments}
\label{sec:torsion_pendulum_experiments}

\subsection{1-D plane-parallel approximation}
\label{subsec:1-D_plane-parallel_approximation}

Thus far we have studied planar configurations because exact solutions exist.  However, torsion pendulum experiments such as \eotwash~do not measure fifth forces in the $\hat z$ direction between planar slabs.  A better approximation is a pair of slabs with features such as holes or grooves.  The $z$ positions of both slabs are fixed, and the source slab is moved in a direction parallel to the planes which we call $\hat x$.  As features in the source mass move past those in the test mass, forces are exerted in the $\pm {\hat x}$ direction.  

\begin{figure}[t]
\begin{center}
\includegraphics[angle=270,width=3.3in]{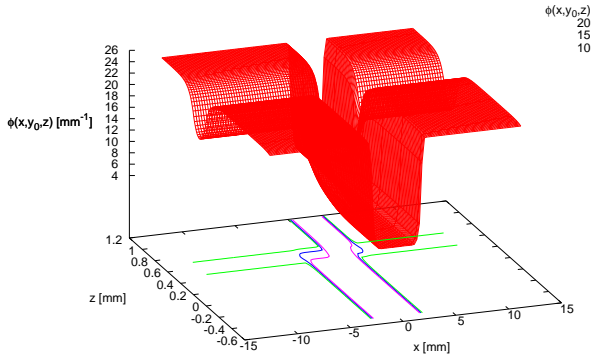}
\caption{Chameleon field for a hole in the source mass moving past another one in the test mass.  The chameleon potential is $V(\phi) = \frac{\lambda}{4!} \phi^4$ with $\lambda=1$, equivalent to (\ref{e:V_chamDE}) with $n=4$ and $\gamma = 1/24$; the matter coupling is $\beta = 1$.  \label{f:3D_numerical}}
\end{center}
\end{figure}

Reference~\cite{Upadhye_Gubser_Khoury_2006} computed the field numerically for such a density configuration.  The field $\phi$ was discretized on a three-dimensional grid of points and the Hamiltonian was minimized with respect to this discrete set of field values.  Figure~\ref{f:3D_numerical} shows the field configuration when the features on the source and test masses are circular holes.  Once the field is known, the force on the test mass, occupying a volume ${\mathcal V}$, can be computed directly from the gradient of the interaction potential, $\vec F = -\int_{\mathcal V} d^3x (\vec\nabla \phi) \rho \Mpl^{-1}$.  However, solving for the field over a range of $x$ positions, for large ranges of $\gamma$, $n$, and $\beta$ values, is computationally expensive.  Accurately accounting for force suppression due to the shielding foil requires discretizing space on length scales $\ll 10$~$\mu$m, yet covering a region of size $\sim 10$~mm.  Thus Ref.~\cite{Adelberger_etal_2007} restricted itself to $n=4$ and $\beta \leq 1$.

\begin{figure}[t]
\begin{center}
\includegraphics[angle=270,width=3.3in]{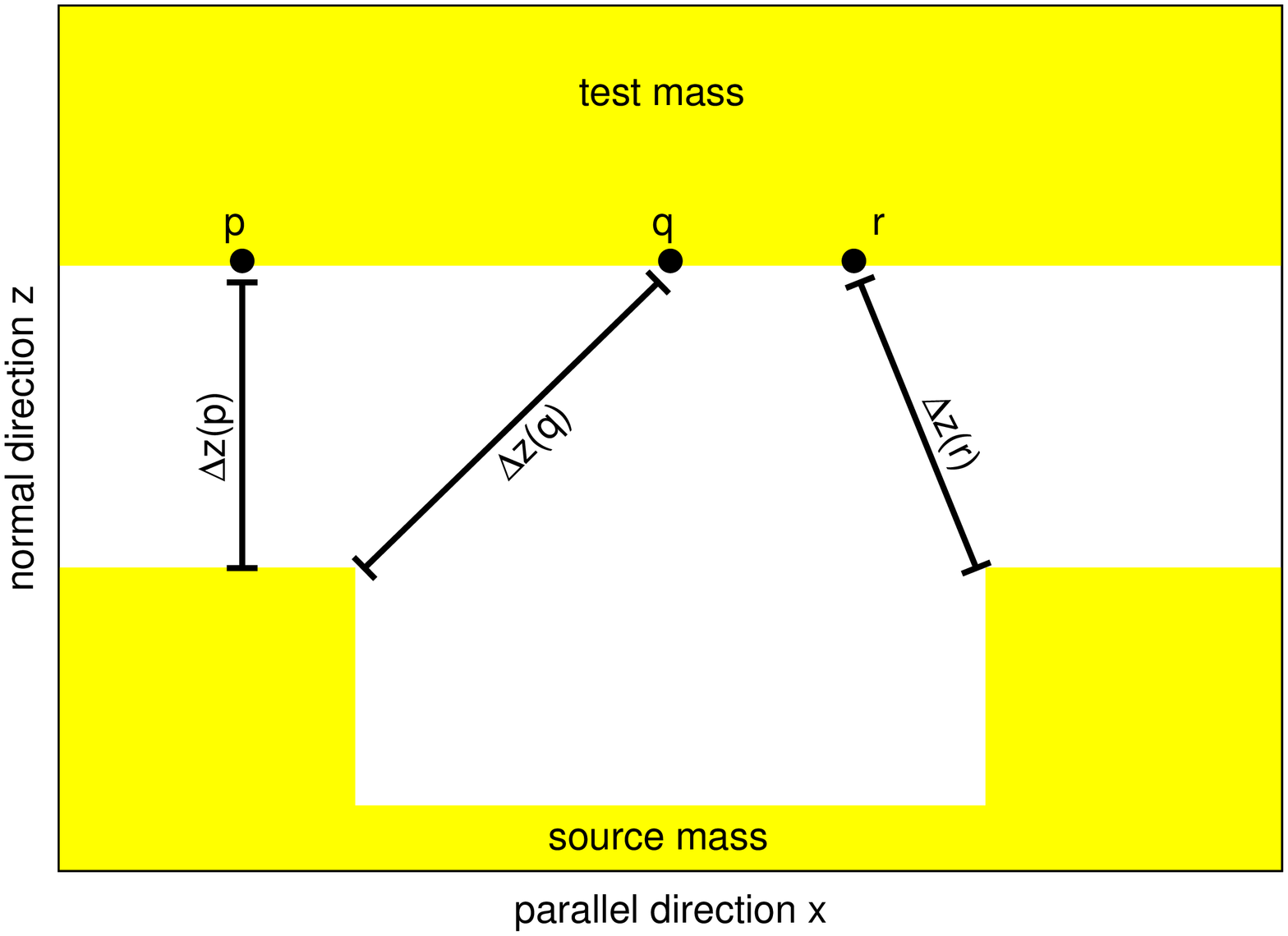}
\includegraphics[angle=270,width=3.3in]{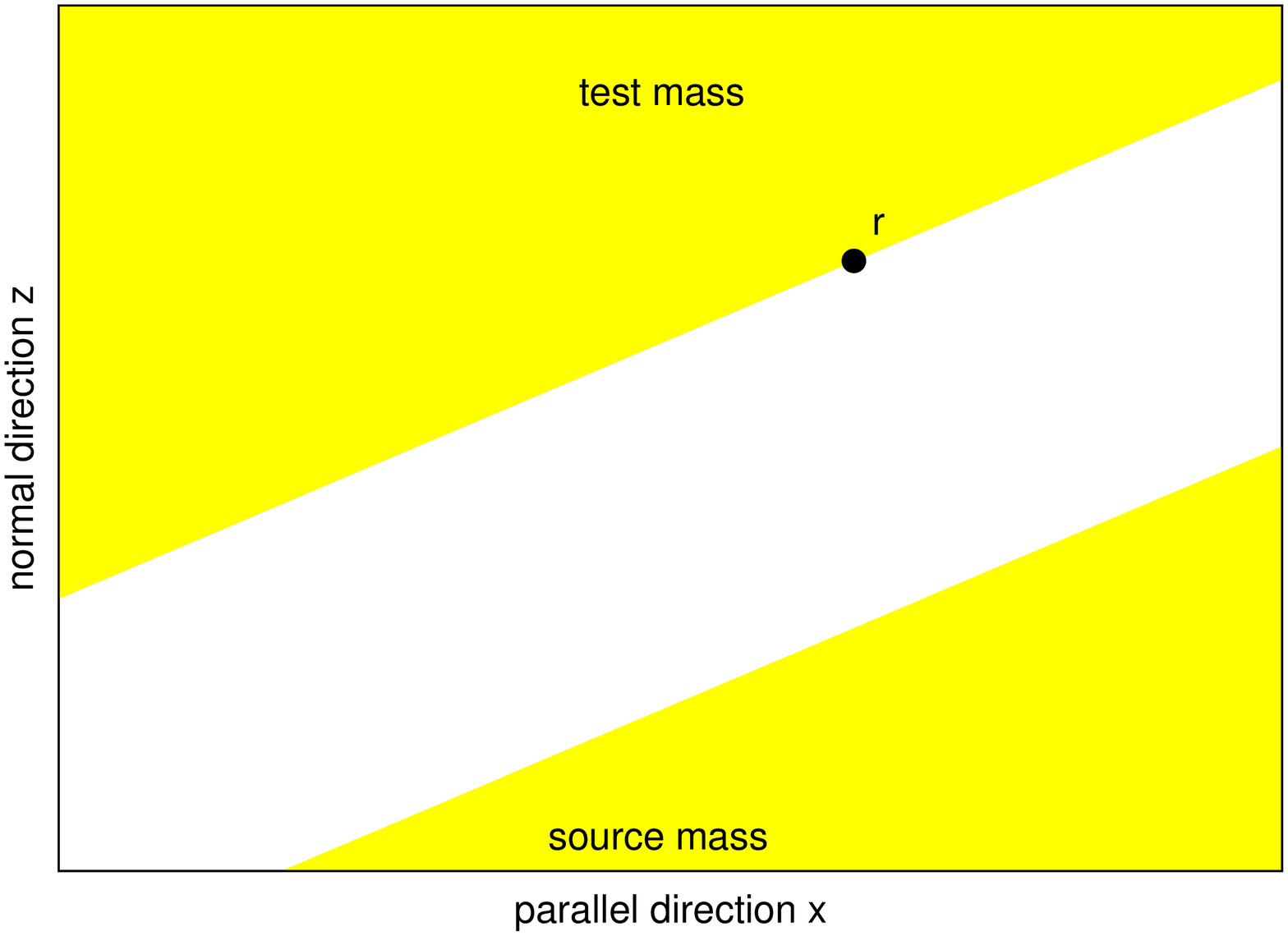}
\caption{One-dimensional plane-parallel approximation.  \Figtop~A feature, such as a groove or hole, on the source mass faces the test mass.  Three points, $p$, $q$, and $r$, are labeled on the test mass, and $\Delta z(p)$, $\Delta z(q)$, and $\Delta z(r)$ are respectively, the distances between each of $p$, $q$, and $r$ and the nearest point on the source slab.  
\Figbot~1Dpp approximation at $r$.   \label{f:1Dpp_approx}}
\end{center}
\end{figure}

Here we make a series of approximations to which we refer collectively as the one-dimensional plane-parallel (1Dpp) approximation.  Consider two parallel slabs as in Figure~\ref{f:1Dpp_approx}~\Figtop, with a hole or groove in the lower (source) slab.  Our goal is to estimate the field at each point on the surface of the source and test masses by approximating the matter distribution near that point as a planar gap.  We use the results of Sec.~\ref{subsec:planar_gap}, which found the surface field $\phir(\Delta z)$ as a function of planar gap size $\Delta z$.  At any point on the surface of either slab, let $\Delta z$ be the distance to the nearest point on the opposite slab, and approxiate the surface field as $\phir(\Delta z)$.  Using this field profile, we compute the energy.  The force in the $\hat x$ direction is the rate of change of this energy as a hole on the source mass passes by one on the test mass.  

Let the distance between the source and test slabs be $\Dzst$, and consider points $p$, $q$, and $r$ on the surface of the test slab, as shown in Fig.~\ref{f:1Dpp_approx}~\Figtop.  Our one-dimensional plane-parallel calculation makes the following approximations.
\begin{enumerate}
\item Each slab is thick enough that the chameleon attains its bulk value deep inside.\label{item:1Dpp_nonlinear}
\item The field at points such as $p$, which are not directly across from the hole on the opposite slab, is equal to the surface field $\phir$ in a one-dimensional planar gap of width $\Delta z(p) = \Dzst$ as studied in Sec.~\ref{subsec:planar_slab_in_vacuum}.  \label{item:1Dpp_overlapping_region}
\item The field at a point such as $q$ or $r$, which is directly across from the hole in the opposite mass, is equal to the surface field $\phir$ in a planar gap with $\Delta z$ equal to the distance to the nearest point on the opposite slab.  For example, $\phi$ at $q$ is equal to the surface field in a gap of size $\Delta z(q)$ shown in the figure. \label{item:1Dpp_non-overlapping_region}
\item Since the surface field should not change on length scales larger than the Compton wavelength, we neglect the energy due to a transition region of width $\meff^{-1}$ at the edge of the region across from a hole.\label{item:1Dpp_neglect_transition_energy}
\item When computing the total energy of a configuration, only the field inside the source and test masses will be counted; changes in the field profile inside holes in each disk, as well as in the gap between disks, are neglected. \label{item:1Dpp_neglect_gap_energy}
\item A shielding foil of thickness $\Delta z_\mathrm{foil}$ between source and test masses reduces the force by a factor of $\fsup = \textrm{sech}(2\meff\Delta z_\mathrm{foil})$ where $\meff$ is the effective mass of the chameleon field at the bulk density of the foil. \label{item:force_suppression}
\end{enumerate}
Figure~\ref{f:1Dpp_approx}~\Figbot~shows this 1Dpp approximation at point $r$.  The geometry of Fig.~\ref{f:1Dpp_approx}~\Figtop~at $r$ is replaced by a one-dimensional planar gap in which the field can be calculated simply.  

Several of these approximations cause us to underestimate the force somewhat.  In particular, the approximation~\ref{item:1Dpp_non-overlapping_region} above adds matter near the opposite mass.  This means that even in non-overlapping regions, the field in the 1Dpp approximation will be closer to its bulk value.  Thus the energy difference as the source mass moves is underestimated, leading to an underestimated force.  Furthermore, aligning holes in the source and test masses will lower the energy associated with the field inside the holes and gaps as well as inside the material of the slabs themselves.  Approximation~\ref{item:1Dpp_neglect_gap_energy} ignores this energy change, leading to an underestimate of the force.  Approximations~\ref{item:1Dpp_nonlinear} and \ref{item:force_suppression} also lead to slight underestimates.

By its nature, the one-dimensional plane-parallel approximation will predict no torque in \eotwash~due to a massless field, hence no sensitivity to a $1/r^2$ force such as Newtonian gravity.  At large $\Dzst$ we can use (\ref{e:phi_vacuum_exterior}) to approximate $\meff \sim \phi^{(n-2)/2} \sim 1/\Dzst$.  Thus the chameleon becomes effectively massless in the limit that $\Dzst$ is much larger than the sizes of the features in the disks.  Since \eotwash~is sensitive to Newtonian torques, the 1Dpp approximation will underestimate the torque in this limit.  Such an underestimate is not significant in \eotwash, whose chameleon constraints are dominated by separation distances $\Dzst$ much smaller than the diameters of the holes.  However, it does mean that the 1Dpp approximation will substantially underestimate the signal in a  ``chameleon lightning-rod'' experiment such as that suggested by~\cite{Jones-Smith_Ferrer_2012}.

On the other hand, approximation~\ref{item:1Dpp_overlapping_region} potentially leads to an underestimate of the energy at points such as $p$, which are not directly across from the hole on the source slab.  If $p$ is within a few Compton wavelengths of the edge of the hole, then the field there will be somewhat larger than expected for a gap of width $\Dzst$, hence its energy will be somewhat greater.  Thus approximation~\ref{item:1Dpp_overlapping_region} increases the energy difference between overlapping and non-overlapping regions, leading to an overestimate of the force.  We shall see in Sections~\ref{subsec:constraints}-\ref{subsec:forecasts} that this overestimate is small for the current-generation \eotwash~experiment but nontrivial for the next-generation experiment.  Additionally, the size of the transition region in approximation~\ref{item:1Dpp_neglect_transition_energy} is just an estimate; it could be $1.5$ or $2$ Compton wavelengths rather than one.  Our choice above will lead to a slight overestimate of the force for the lowest chameleon couplings.

\subsection{Torsion pendulum}
\label{subsec:torsion_pendulum}

\begin{figure}[t]
\begin{center}
\includegraphics[width=1.65in]{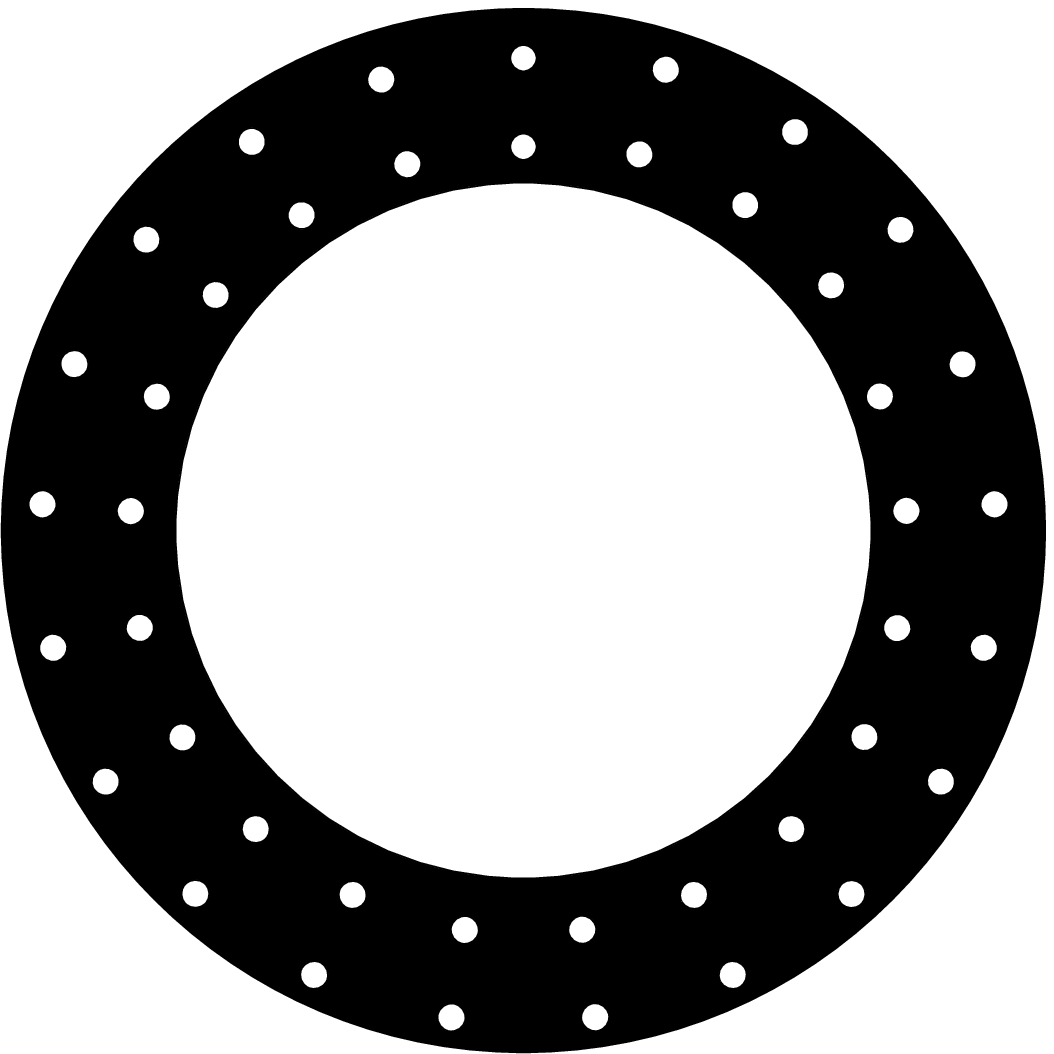}
\includegraphics[width=1.65in]{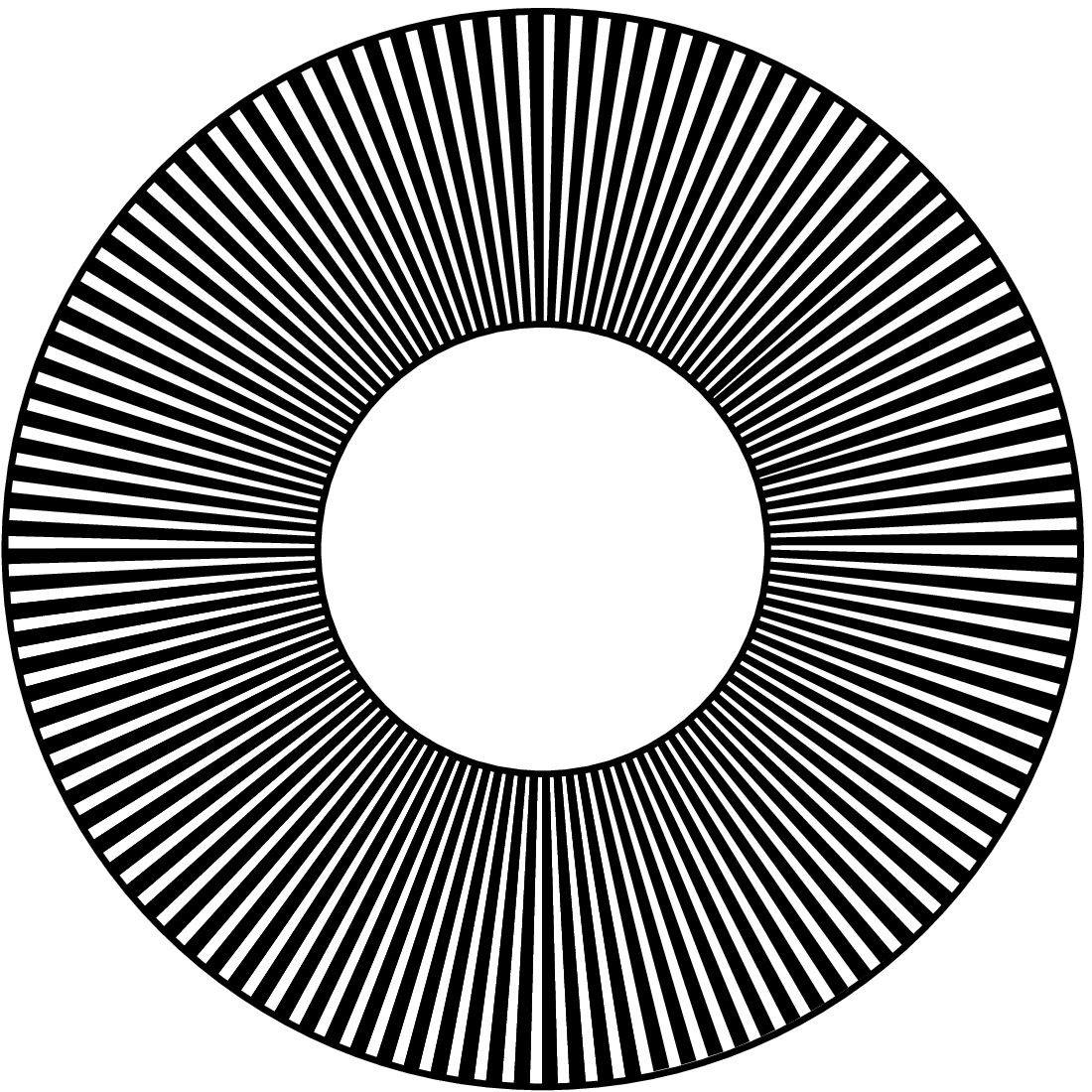}
\caption{Geometry of disks used in \eotwash~torsion pendulum experiment (from~\cite{Upadhye_Gubser_Khoury_2006}; not to scale).  \Figleft~Current experiment, with two rows of 21 holes in each disk. \Figright~Next-generation experiment, with 120 radial grooves in each disk.  \label{f:eotwash_disks}}
\end{center}
\end{figure}

Here we apply the 1Dpp approximation of the previous section to a hypothetical torsion pendulum similar to the \eotwash~experiment \cite{Kapner_etal_2007}.  Such an experiment consists of a pair of parallel, rotating disks with matching holes at regular intervals, as in Figure~\ref{f:eotwash_disks}.  The lower disk, the source mass or ``attractor,'' is mounted on a turntable which keeps it rotating uniformly.  The upper disk, the test mass or ``detector,'' is a torsion pendulum allowed to rotate freely.  If there is a fifth force, then the test mass will experience small torques as the holes on the source mass move in and out of alignment with those on the test mass.  

The chameleon fifth force which results from moving a hole on the source disk past one on the test disk is the energy cost per unit distance of the change in the field configuration.  Using the 1Dpp approximation, we can estimate the field configuration on the surface of each disk as one hole of radius $\rsh$ on the source disk rotates past another of radius $\rth$ on the test disk.  Assume that both disks have the same density $\rhom$.  Let the origin of the coordinate system be the point on the source disk directly across from the center of the test disk hole, with $\hat z$ parallel to the rotation axis of the disks and $\hat x$ in the direction of motion of the source hole (that is, the tangential direction).  If the source hole is far away, then the field at a position $r = \sqrt{x^2 + y^2}$ on the surface $z=0$ of the source disk is the surface field in a gap of size $\Delta z = \sqrt{(\Dzst)^2 + (\rth-r)^2}$.  Since $\Delta z > \Dzst$, the field $\phi$ will be greater on the portion of the source disk across from the test disk hole.  Thus $\phi$ will be farther away from its energy-minimizing value $\phiB(\rhom)$.  On the other hand, if the source and test holes overlap, then $\Delta z = \Dzst$ over the maximum possible area on both disks, minimizing the energy.

\begin{figure}[t]
\begin{center}
\includegraphics[angle=270,width=3.3in]{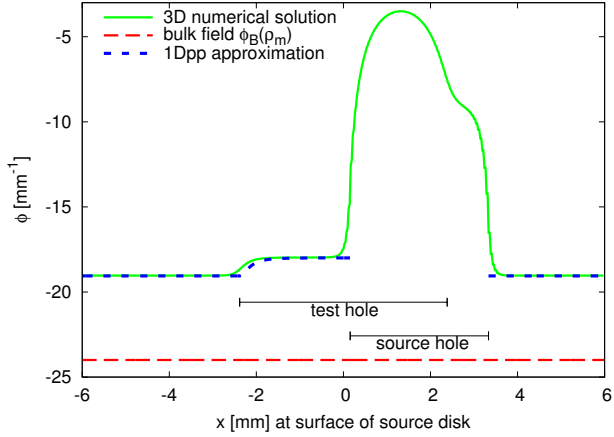}
\caption{Field on the surface of the source mass, with source and test holes offset, for $V(\phi) = \frac{\lambda}{4!}\phi^4$, $\lambda = 1$, $\beta = 1$, and $\Dzst = 0.2$~mm.  The horizontal axis shows the tangential direction, and $x=0$ coincides with the center of the test mass hole.  The 1Dpp approximation agrees quite well with the 3D numerical calculation of~\cite{Upadhye_Gubser_Khoury_2006}.  \label{f:source_surface}}
\end{center}
\end{figure}

\begin{figure}[tb]
\begin{center}
\includegraphics[angle=270,width=1.65in]{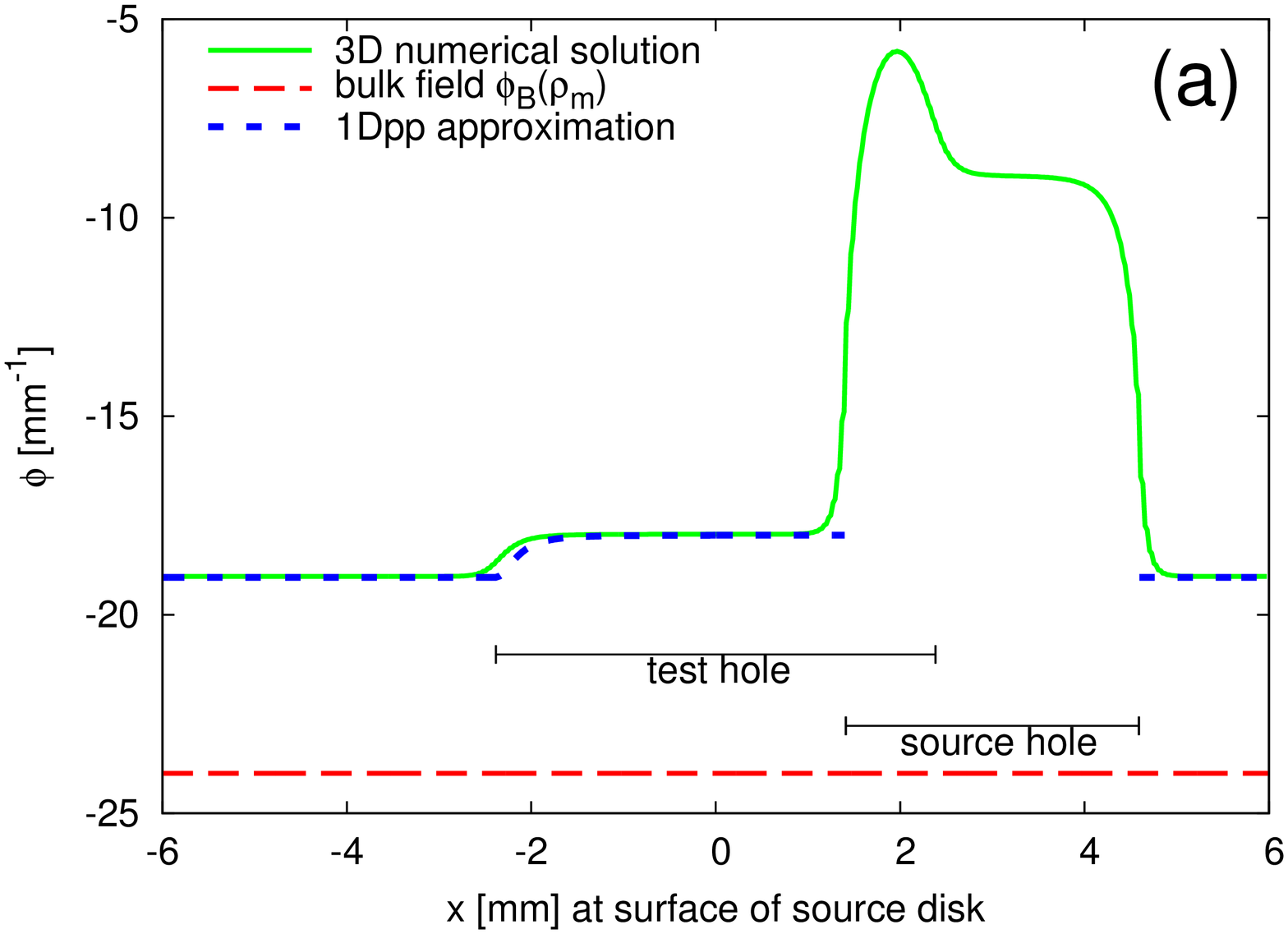}
\includegraphics[angle=270,width=1.65in]{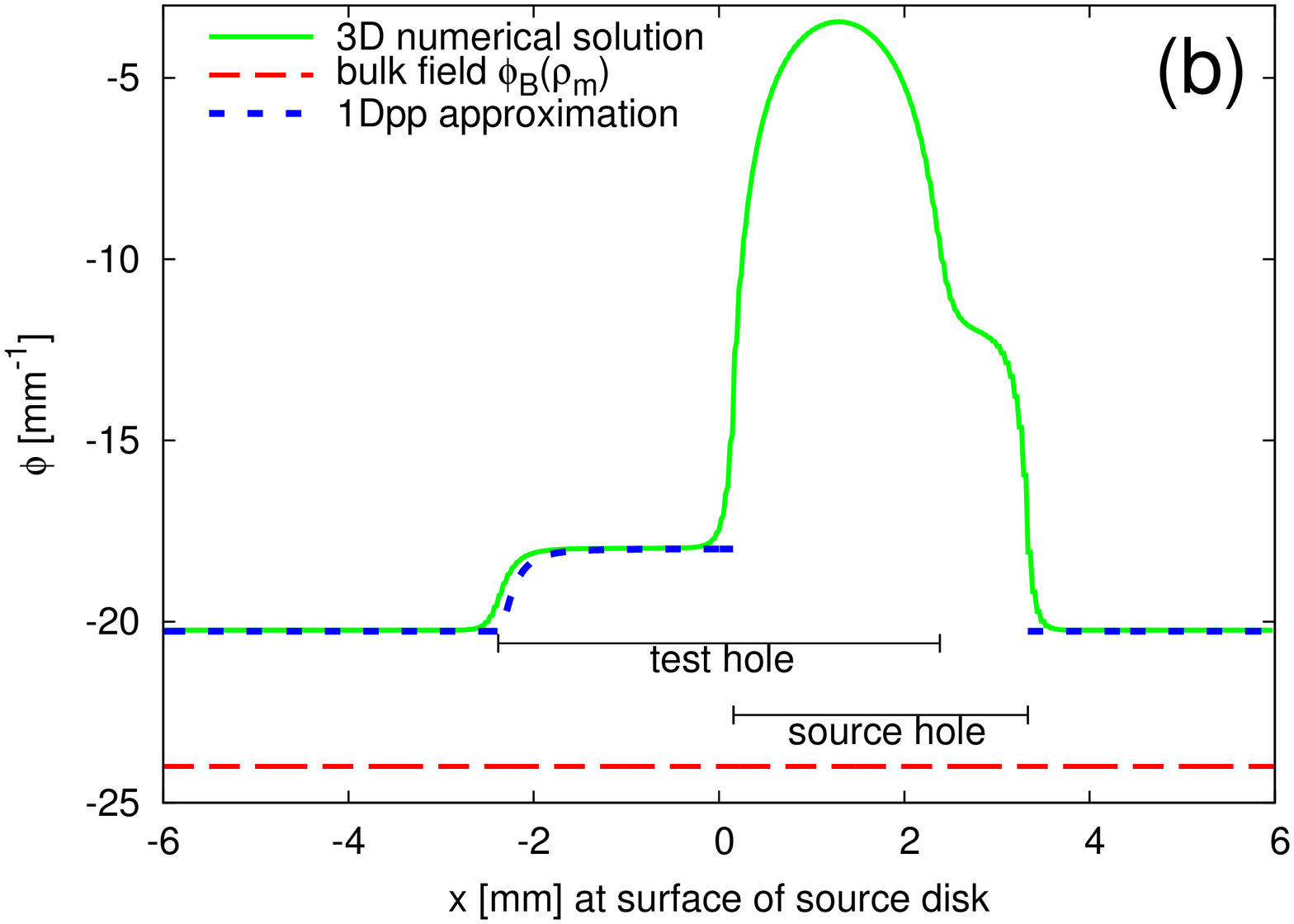}
\includegraphics[angle=270,width=1.65in]{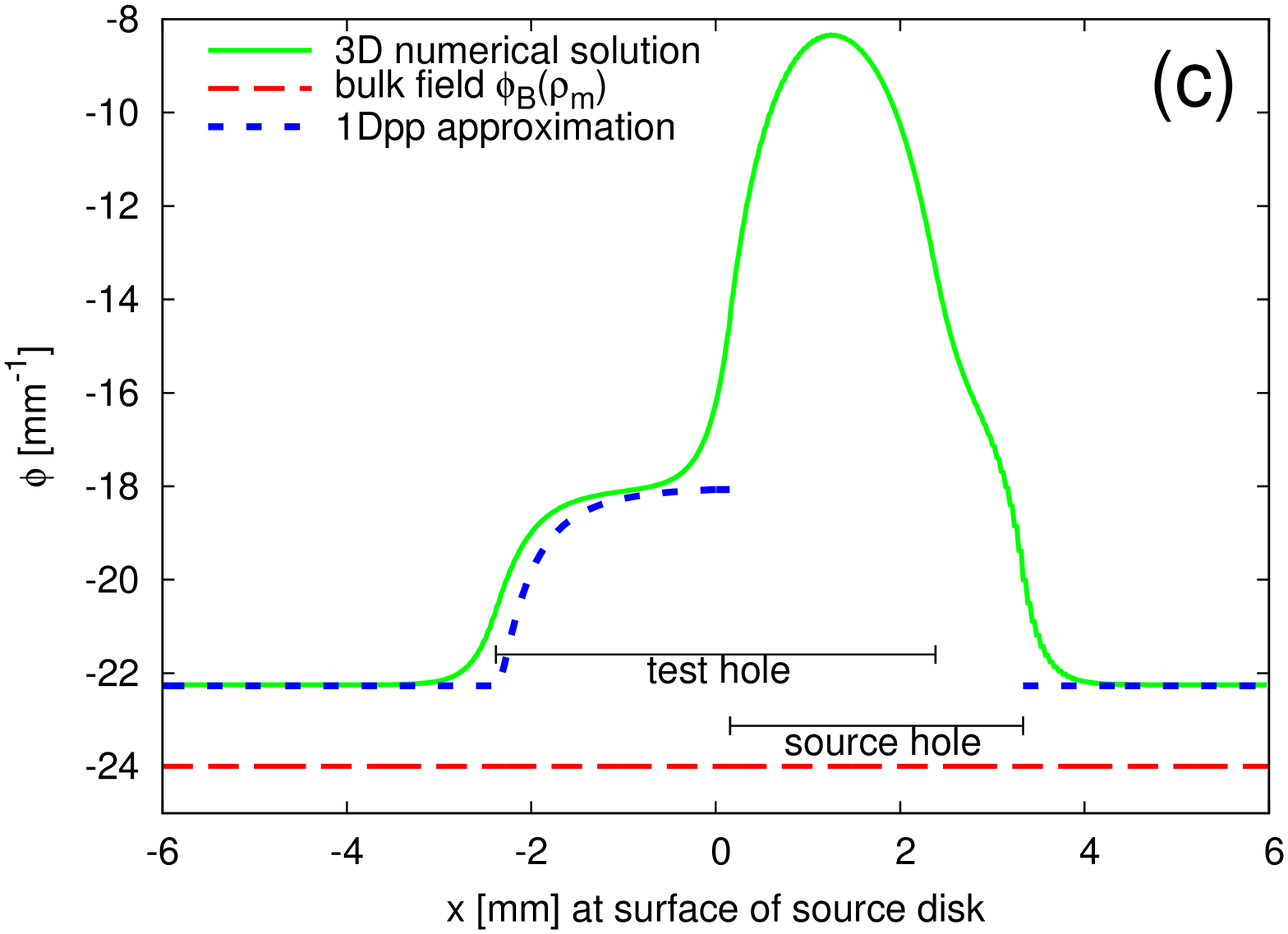}
\includegraphics[angle=270,width=1.65in]{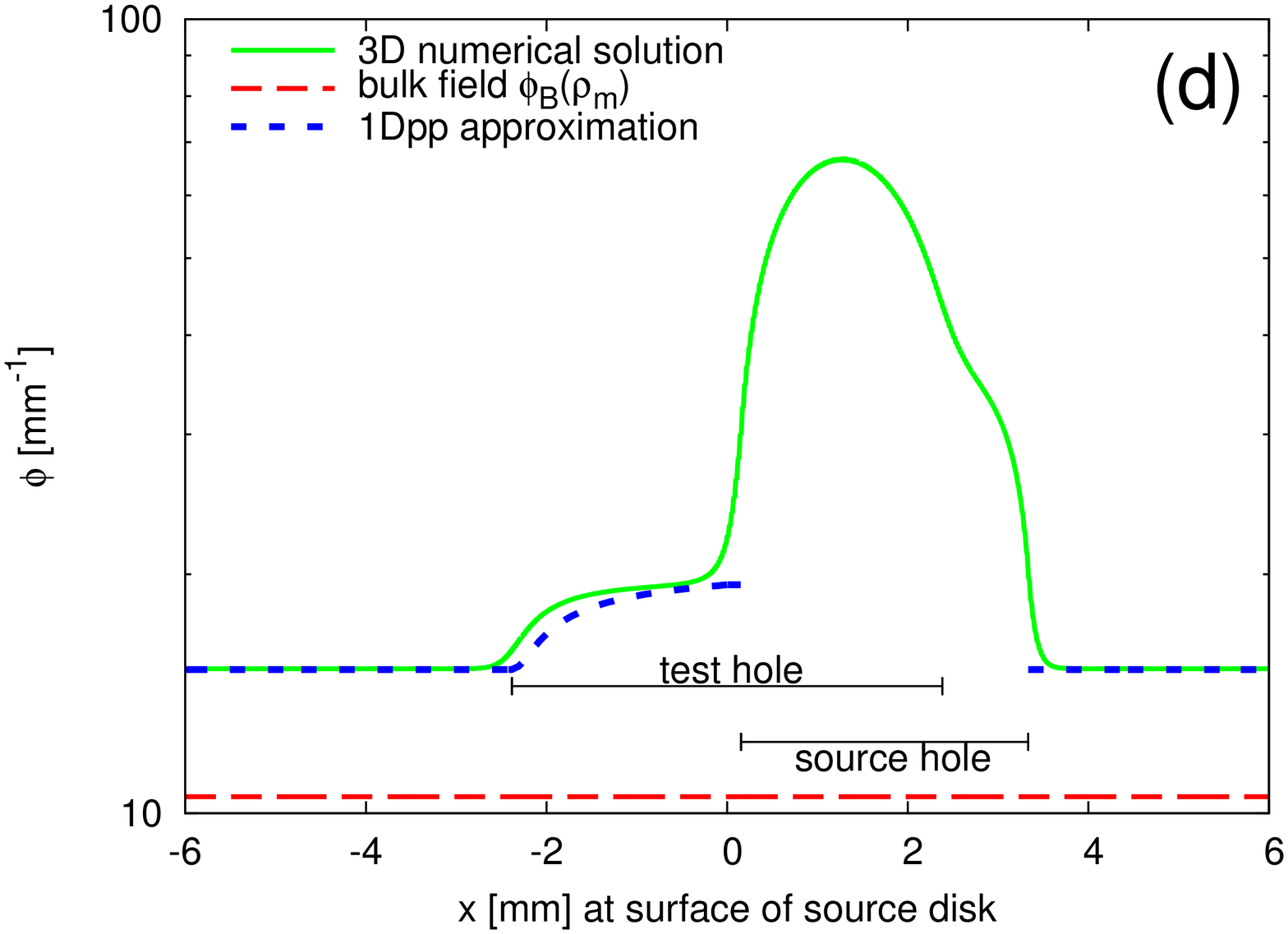}
\caption{Similar to Fig.~\ref{f:source_surface}, but with \Figa~a greater source hole displacement, \Figb~$\Dzst = 0.1$~mm, \Figc~$\lambda = \beta = 0.1$, and \Figd~$n=-1$, $\gamma = 1$, $\beta = 1$.\label{f:source_surface2}}
\end{center}
\end{figure}

Figure~\ref{f:source_surface} shows the field profile $\phi(x,0,0)$ on the surface of the source disk when the source hole is displaced from the test hole.  The 1Dpp approximation is in close agreement with the three-dimensional numerical calculation of~\cite{Upadhye_Gubser_Khoury_2006} except for a transition region at the edge of the hole.  Figure~\ref{f:source_surface2} compares the 1Dpp approximation and the 3D numerical calculation for a range of geometries and models.  In all cases the two agree reasonably well.

Now that the 1Dpp approximation has given us the field $\phis(x,y,0)$ on the surface of the source disk, we may find the energy.  Assuming that the disk is a thick slab, we approximate the field inside it using the thick-slab linearization~(\ref{e:thick-slab_linearization}).  Then the energy inside the region of the source disk across from the test mass hole, assuming that the source hole is far away, is
\begin{eqnarray}
\Eh
&=&
\int_0^{\rth} 2\pi r \,dr \int_0^{-\infty}dz 
\left(
\frac{1}{2}\left|\vec\nabla \phi\right|^2 + V(\phi)
\right)
\nonumber\\
&=&
\int_0^{\rth} \frac{\pi r \, dr}{2\meff}
\left[
2\meff^2 (\phis - \phiB)^2 e^{2 \meff z}
+
\left|\frac{\partial\phi}{\partial r}\right|^2
\right]\qquad
\label{e:E_hole}
\end{eqnarray}
where $\phiB$ and $\meff$ are evaluated at the disk density $\rhom$.  The subscript h denotes the region of the source disk across from the hole.

Next, consider a region of the source disk far from any hole on the test disk.  In that region, the two disks will look like a pair of parallel planes with a separation $\Dzst$, so $\phis$ will be a constant on the surface.  Let $\Enh$ be the energy of a region of the same size, where the subscript nh is short for ``no hole.''  Then $\DEs = \Eh - \Enh$ is the energy cost in the source disk associated with each hole in the test disk.  Similarly, we may compute $\DEt$, the energy cost in the test disk.

\begin{figure}[t]
\begin{center}
\includegraphics[angle=270,width=3.3in]{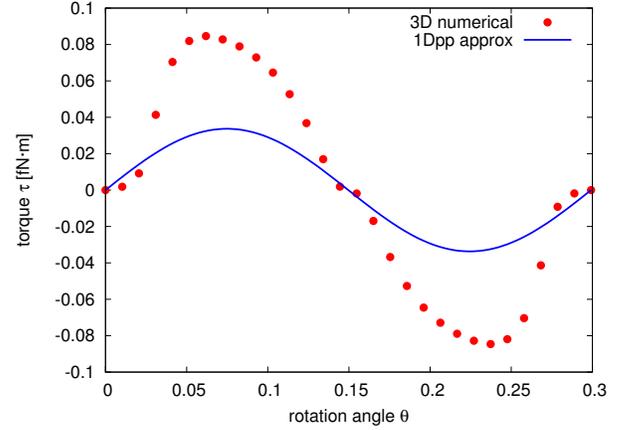}
\caption{Comparison between the 1Dpp approximation and the 3D numerical computation of~\cite{Upadhye_Gubser_Khoury_2006} for a $\phi^4$ chameleon with $\lambda = \beta = 1$ and disk separation $\Dzst = 0.1$~mm.  \label{f:comparison_3D_1Dpp}}
\end{center}
\end{figure}

Finally, we may compute the total energy and torque.  The amplitude of the total energy $\Etot$ will be half of the total energy change $\DEs+\DEt$, multiplied by the total number $\Nholes$ of holes, which is $42$ for \eotwash.  Let $\theta$ be the rotation angle, and define $\theta=0$ to be the angle at which source and test disk holes are perfectly overlapping.  For equally-spaced holes in $\Nrows=2$ rows, the frequency with which $\Etot$ varies is $\omegah = \Nholes/\Nrows$.  Multiplying by the force suppresion factor $\fsup$, we obtain the total energy and torque,
\begin{eqnarray}
\Etot
&=&
-\frac{1}{2}\Nholes \fsup (\DEs+\DEt) \cos(\omegah \theta)
\label{e:eotwash_Etot}
\\
\tau
&=&
\frac{1}{2} \Nholes \omegah \fsup (\DEs+\DEt)\sin(\omegah \theta).
\label{e:eotwash_tau}
\end{eqnarray}

\begin{figure}[t]
\begin{center}
\includegraphics[angle=270,width=3.3in]{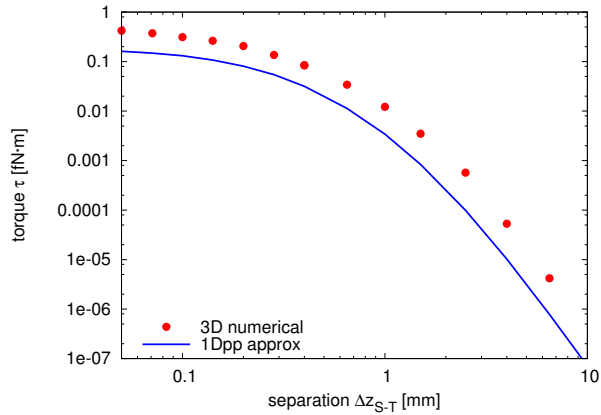}
\caption{Comparison between the 1Dpp approximation and the 3D numerical computation of~\cite{Upadhye_Gubser_Khoury_2006} for a $\phi^4$ chameleon with $\lambda = 1/10$ and $\beta = 1$, as a function of $\Dzst$.  \label{f:comparison_vary_Dzst}}
\end{center}
\end{figure}

Figure~\ref{f:comparison_3D_1Dpp} compares (\ref{e:eotwash_tau}) to the three-dimensional numerical computation of~\cite{Upadhye_Gubser_Khoury_2006}.  The 1Dpp approximation underestimates the torque by a factor of about $2.5$.   Figure~\ref{f:comparison_vary_Dzst} demonstrates that this underestimate becomes worse by a factor of about two at separations around a few millimeters, the diameters of the source and test holes, as expected from Sec.~\ref{subsec:1-D_plane-parallel_approximation}.  However, since the torque itself falls off rapidly with separation distance, constraints will be dominated by small $\Dzst$.  Thus this worsening of the 1Dpp approximation at large $\Dzst$ will not have a significant effect on the final constraints. 

\subsection{Constraints}
\label{subsec:constraints}

Using the 1Dpp approximation developed above, we may quickly estimate constraints on power law chameleon models from the current-generation \eotwash~experiment.  This experiment has $\Nholes=42$, $\Nrows=2$, $\rth = 2.4$~mm, and $\rsh = 1.6$~mm.  The source and test disks were made of molybdenum, with a density $\rhom = 10$~g/cm$^3$, while the laboratory vacuum density was $\rhov = 10^{-6}$~torr $\sim 10^{-12}$~g/cm$^3$.  \eotwash~probes torques over a range of disk separations; however, here we approximate the experiment as excluding torques greater than $0.003$~fN$\cdot$m at $\Dzst = 0.1$~mm.

\begin{figure}[t]
\begin{center}
\includegraphics[angle=270,width=3.3in]{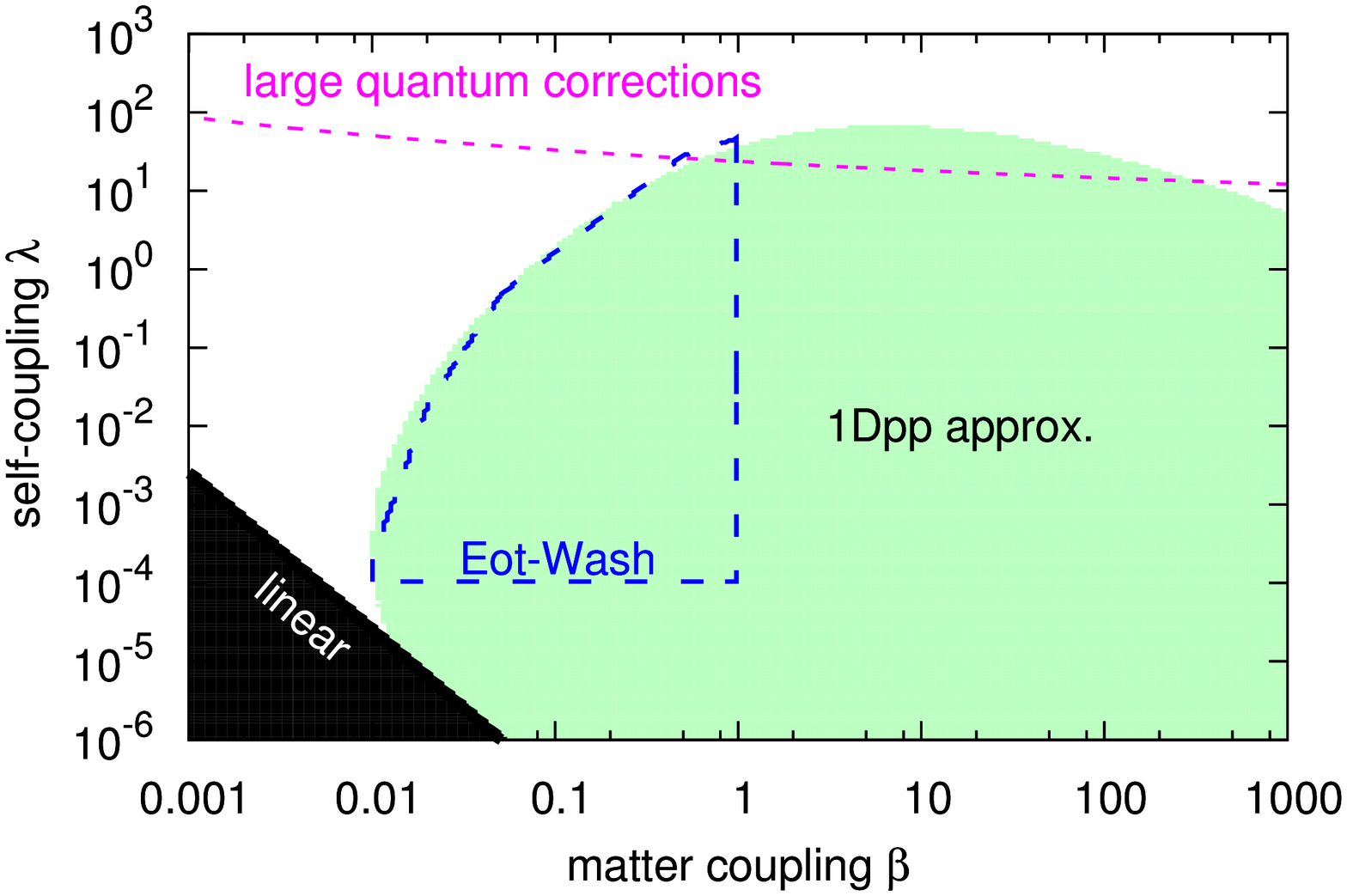}
\includegraphics[angle=270,width=3.3in]{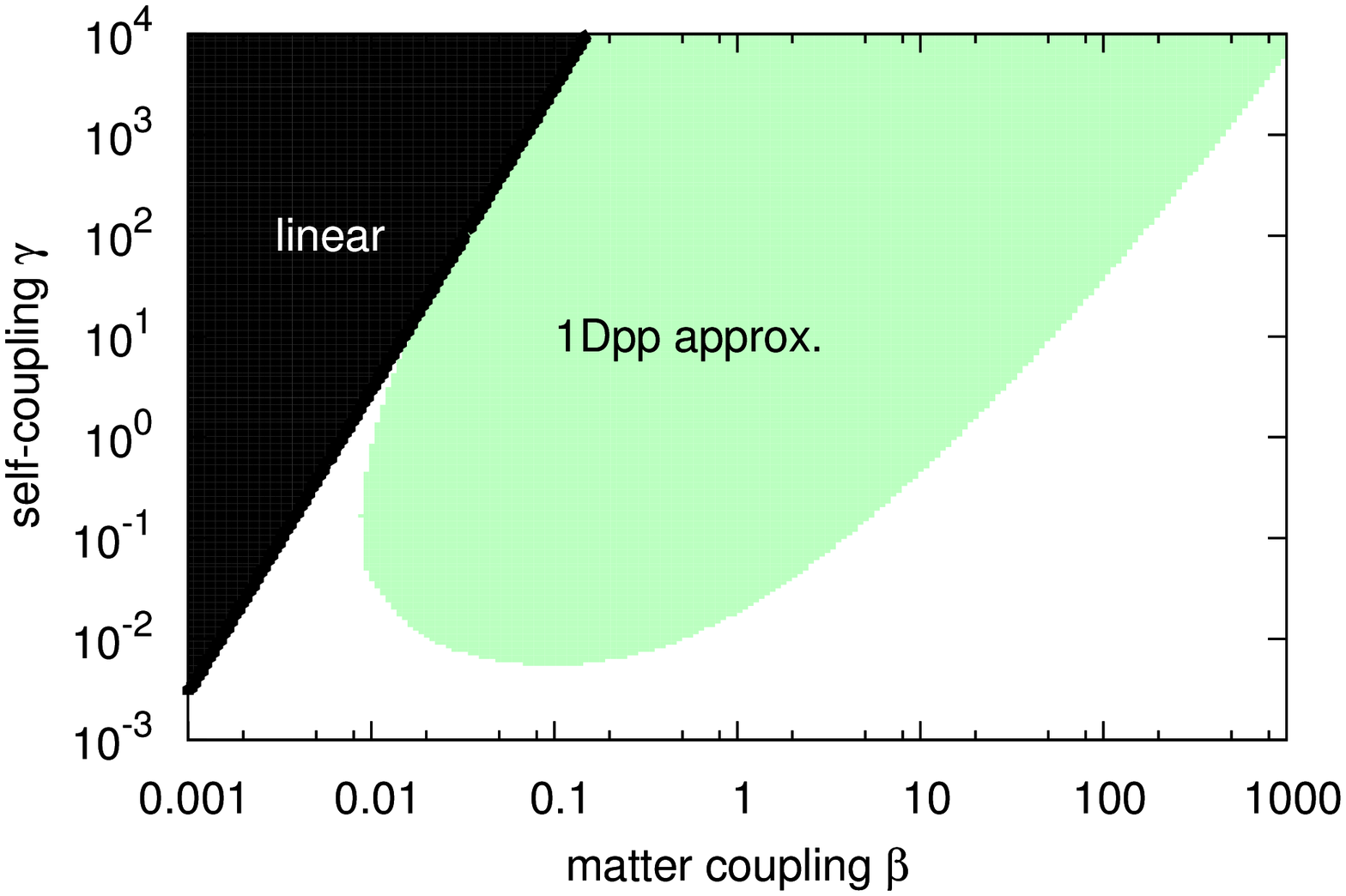}
\caption{1Dpp excluded regions (shaded light green).   The black  shaded regions identify models which are linear inside the source and test masses; these are excluded unless $\beta$ is very small.  
\Figtop~Quartic chameleon, $V(\phi) = \frac{\lambda}{4!}\phi^4$.  The long-dashed blue line shows the constraints of~\cite{Kapner_etal_2007,Adelberger_etal_2007}.  Models above the short-dashed purple line have large quantum corrections.
\Figbot~Inverse power law chameleon, $V(\phi) = M_\Lambda^4(1 + \gamma M_\Lambda/\phi)$.  All models shown have small quantum corrections.
\label{f:constraints_n4_n-1}}
\end{center}
\end{figure}

Figure~\ref{f:constraints_n4_n-1} shows our approximate 1Dpp \eotwash~constraints.  In particular, Fig.~\ref{f:constraints_n4_n-1}~\Figtop~compares 1Dpp constraints to the more precise numerical calculation of~\cite{Adelberger_etal_2007} for the $\phi^4$ chameleon.  In the range $0.01 \leq \beta \leq 1$ covered by both sets of constraints, the 1Dpp exclusion lower bound on $\lambda$ agrees well with the more precise calculation.  

The 1Dpp calculation underestimates constraints near $\beta=1$ due to Approxmation~\ref{item:1Dpp_neglect_gap_energy} of Sec.~\ref{subsec:1-D_plane-parallel_approximation}, which neglects the contribution to the total energy of the field in the gap between disks.  We could potentially correct for this underestimate by including an extra factor in (\ref{e:eotwash_Etot},~\ref{e:eotwash_tau}) and using the numerical computations of~\cite{Upadhye_Gubser_Khoury_2006} to calibrate this factor. Meanwhile, around $\beta = 0.01$, the 1Dpp calculation overestmates constraints.  Approximation~\ref{item:1Dpp_neglect_transition_energy} of Sec.~\ref{subsec:1-D_plane-parallel_approximation} assumes a transition region of width $\meff^{-1}$ associated with each test and source mass hole, but this is just an estimate. We could potentially include another factor parameterizing the number of Compton wavelengths in the transition region, and then adjust it to match Ref.~\cite{Upadhye_Gubser_Khoury_2006} more closely.  

Since the goal of the present work is an estimate of \eotwash~constraints rather than a rigorous data analysis, we do not fit these two ``fudge factors'' to~\cite{Upadhye_Gubser_Khoury_2006}.  We have chosen an \eotwash~exclusion limit of $0.003$~fN$\cdot$m such that the 1Dpp constraints approximately match those of~\cite{Adelberger_etal_2007}, which is equivalent to estimating a value for the first of these factors.  Our choice is roughly consistent with the \eotwash~data presented in~\cite{Kapner_etal_2007}, and the resulting 1Dpp constraints are a slight underestimate  in the strongly nonlinear regime $\beta \gtrsim 1$.  Meanwhile, we do not adjust the second factor at all.

Models in Fig~\ref{f:constraints_n4_n-1}~\Figtop~above the dashed purple line have large quantum corrections; they fail the quantum stability conditions discussed in Sec.~\ref{subsec:quantum_stability_condition}.  For a range of matter couplings $1 \lesssim \beta \lesssim 100$, \eotwash~excludes all quantum-stable $n=4$ chameleon models.  

Constraints on the $n=-1$ chameleon are shown in Figure~\ref{f:constraints_n4_n-1}~\Figbot.  A more rigorous analysis such as~\cite{Adelberger_etal_2007} does not exist for this model.  The quantum stability conditions do not exclude any models shown here; quantum corrections are $\approx 20\%$ of the tree level values in the bottom right corner of the plot, and smaller elsewhere.  For $\gamma = 1$, the model in which the power law term in the potential~(\ref{e:V_chamDE}) has the same energy scale as the dark energy, \eotwash~excludes $0.01 < \beta < 15$ in the 1Dpp approximation.

\begin{figure}[t]
\begin{center}
\includegraphics[angle=270,width=3.3in]{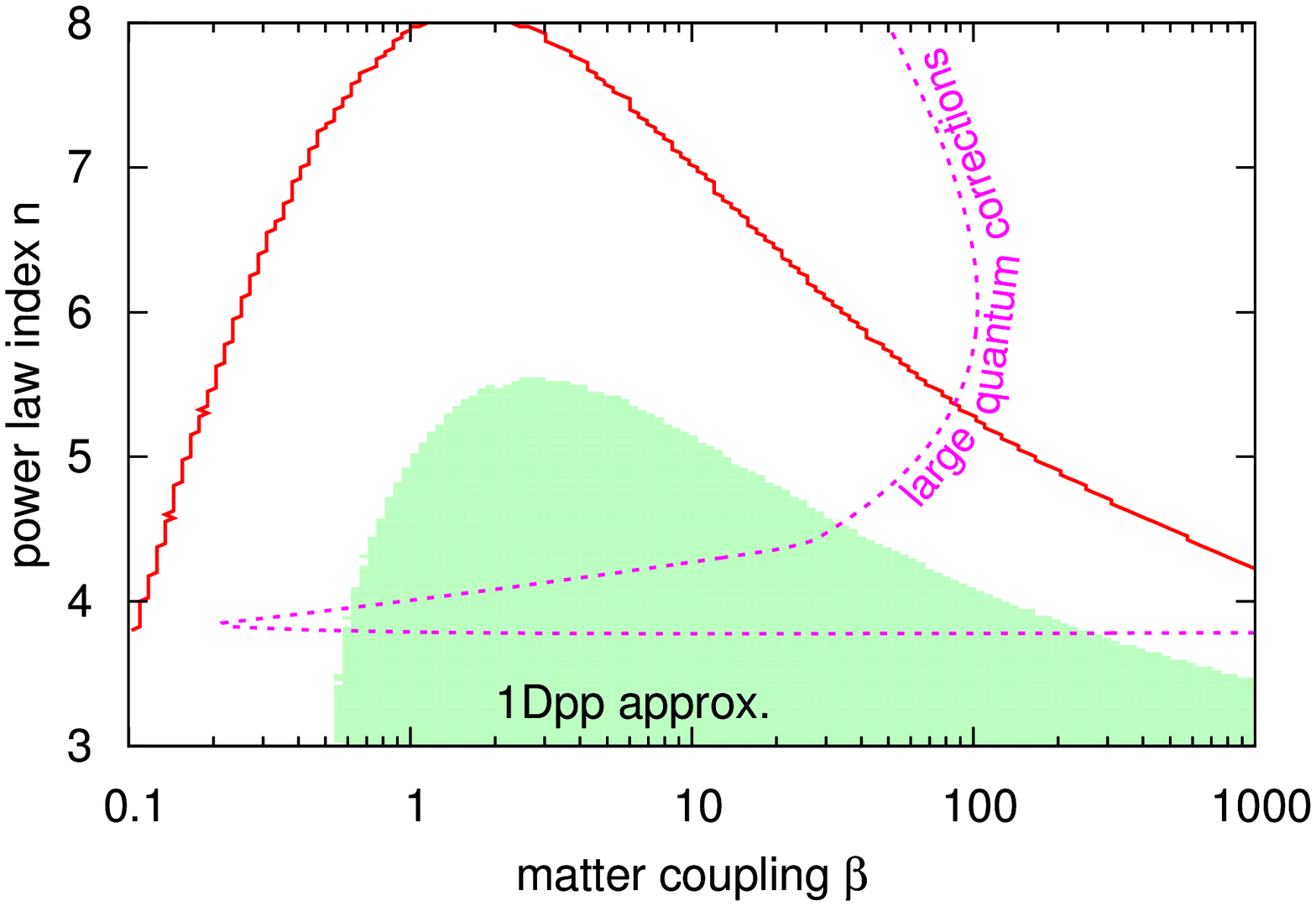}
\includegraphics[angle=270,width=3.3in]{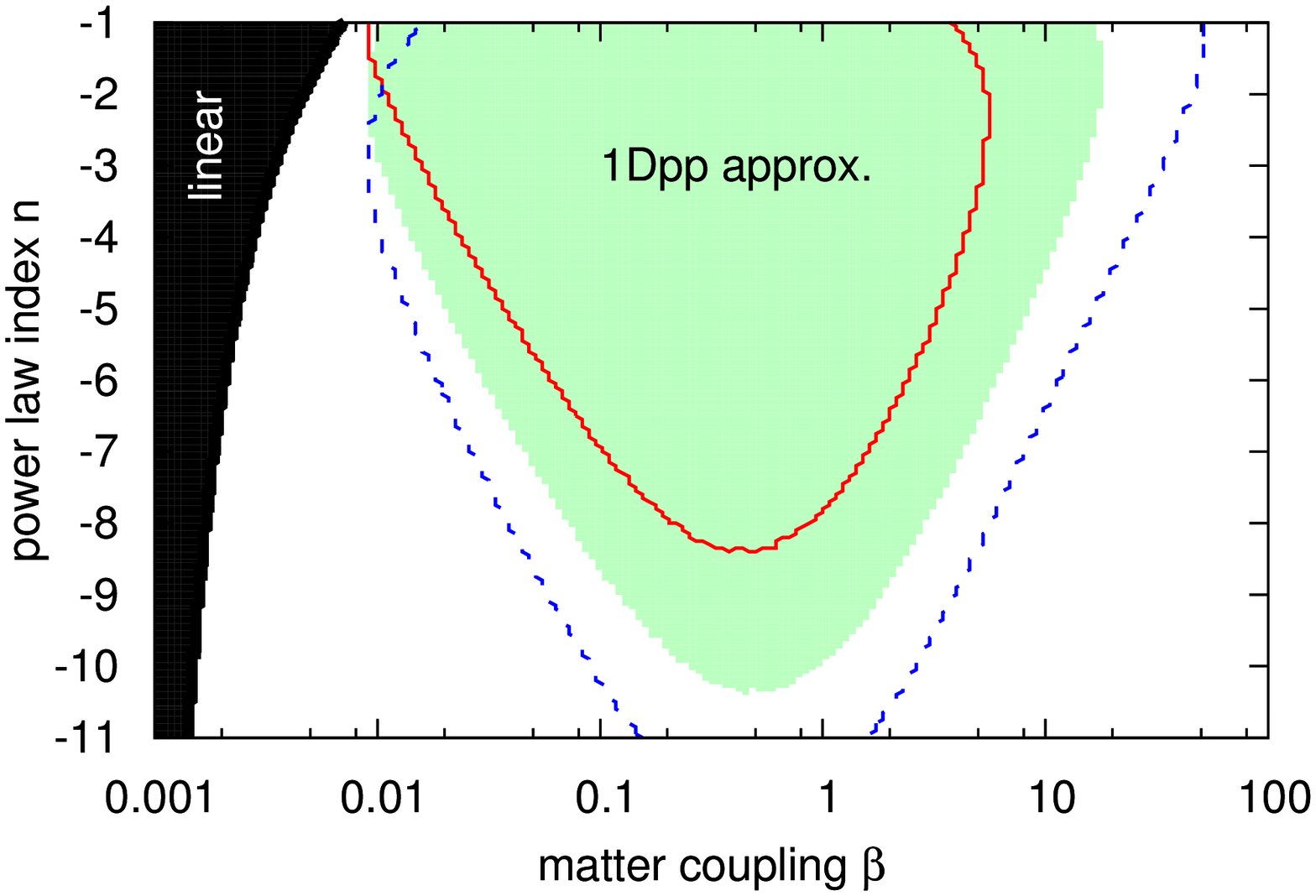}
\caption{Excluded region in the $\beta,n$ plane, for $\gamma=1$ (green shaded region), $\gamma=0.1$ (solid red line), and $\gamma=10$ (dashed blue line).  
\Figtop~Positive $n$. For $\gamma=1$, models to the right of the purple dashed line have large quantum corrections; in the $\gamma=0.1$ case, all models shown pass the quantum stability test.   For $\gamma=10$ and $n \geq 3$ no models are excluded. 
\Figbot~Negative $n$. All models shown pass the quantum stability test.   
\label{f:constraints_gamma_const}}
\end{center}
\end{figure}

Figure~\ref{f:constraints_gamma_const} shows constraints  in the $\beta,n$ plane for several $\gamma$.  For $n=4$, the self-coupling $\gamma=10$ corresponds to $\lambda=240$, a rather large number for which the chameleon effect is very strong.  Thus there are no constraints for this value in Fig.~\ref{f:constraints_gamma_const}~\Figtop.  

Both plots in Fig.~\ref{f:constraints_gamma_const} show that constraints vanish at large $|n|$.  We can see why by computing the maximum possible force per unit area between two planar slabs of density $\rhom$.  If the distance separating them is small, then the field at the center of the gap will be $\phig \approx \phiB(\rhom)$.  If each slab is sufficiently thick, then the field on the side facing away from the other slab will be $\phisv = \phiB(\rhom)(1-1/n)$.  Then the magnitude of the attractive force between them is $F = \beta \rhom |\phiB(\rhom)-\phisv|/\Mpl = (\beta \rhom/\Mpl) |\phiB(\rhom)/n| \rightarrow \beta \rhom M_\Lambda / |n\Mpl|$ at large n.  Suppose that we also include a force suppression factor $\textrm{sech}(2\meff\Delta z_\mathrm{foil})$.  At large $|n|$, $\meff \sim \sqrt{\beta|n|}$, so the suppression factor decreases quickly.  Thus large-$|n|$ models will be difficult to exclude.

As a final note, we have used the 1Dpp approximation to study the effects on these constraints of a degraded laboratory vacuum.  We find that the chameleon fifth force at $\Dzst = 0.1$~mm is extremely insensitive to the vacuum quality; even conducting the experiment at atmospheric pressure does not noticably reduce the chameleon fifth force.  At $\Dzst = 10$~mm, the largest disk separation probed by \eotwash, constraints at atmospheric pressure are $\sim 10\%$ worse than those in a vacuum for $n=-1$.  

\subsection{Forecasts}
\label{subsec:forecasts}

The geometry of the next-generation \eotwash~source and test disks is shown in Figure~\ref{f:eotwash_disks}~\Figright.  In order to visualize such a disk, one can imagine a pie cut into $240$ equal wedges, with every other wedge removed, and a circular region excised from the center.  We approximate each disk as having an inner radius of $13$~mm and an outer radius of $23$~mm.  Each of the $\Nwedges=120$ wedges has a thickness $\Dzw = 50$~$\mu$m and a height $\Dyw = 10$~mm.  As in~\cite{Upadhye_Gubser_Khoury_2006}, we approximate each wedge as a rectangular sheet of width $\Dxw = 2\pi r_\mathrm{avg} / (2\Nwedges) = 0.47$~mm, where $r_\mathrm{avg} = 18$~mm is the average of the inner and outer radii.  Each wedge has a density $\rhom = 20$~g/cm$^3$, and they are mounted on a glass disk of density $2$~g/cm$^3$.

\begin{figure}[t]
\begin{center}
\includegraphics[angle=270,width=3.3in]{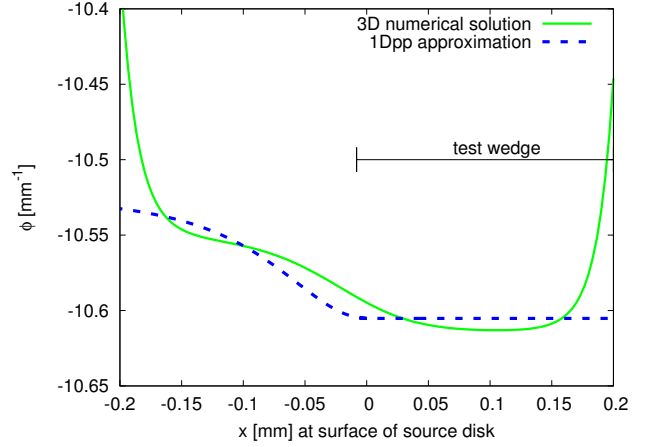}
\caption{Field on the surface of a wedge on the source disk, for $\lambda = 100$, $\beta = 10$, and $\Dzst = 0.1$~mm. \label{f:source_wedge}}
\end{center}
\end{figure}

Figure~\ref{f:source_wedge} shows this 1Dpp approximation along with the more accurate three-dimensional numerical simulation.  Agreement between the two is not as close as it was in Figs.~\ref{f:source_surface}-\ref{f:source_surface2}.  This is because edge effects are larger when the features in the disks are long, narrow grooves rather than circular holes.  However, our approximation reproduces the qualitative features of the field, and particularly the difference in the surface field between regions which do and do not overlap a wedge on the opposite disk.  This field difference determines the energy difference, hence the predicted torque.

Let the gap between source and test disks be $\Dzst$.  We can immediately apply the 1Dpp approximation to determine the energy associated with the overlap between wedges on opposite disks.
\begin{eqnarray}
E
&=&
\int dx \, dy \, dz
\left[
  \frac{1}{2} \left|\vec \nabla \phi\right|^2 + V(\phi)
\right]
\nonumber\\
&=&
\frac{\Dyw f_z}
     {2\meff}
\int_0^\frac{\Dxw}{2} dx
\left[
  2\meff^2 (\phis-\phiB)^2
  + \left|\frac{d\phis}{dx}\right|^2
\right]
\quad
\label{e:E_wedge}
\end{eqnarray}
where $f_z = 1 - \exp(-\meff \Dzw)$ corrects for the finite wedge thickness; as in (\ref{e:E_hole}), $\meff$ and $\phiB$ are evaluated at the bulk density $\rhom$.  In the case of perfect overlap, $\phis(x)$ is a constant equal to the surface field in a gap of size $\Dzst$.  In the case of no overlap, $\phis(x)$ is the surface field in a gap of size $\sqrt{(\Dzst)^2 + (\Dxw/2 - x)^2}$.  

\begin{figure}[t]
\begin{center}
\includegraphics[angle=270,width=3.3in]{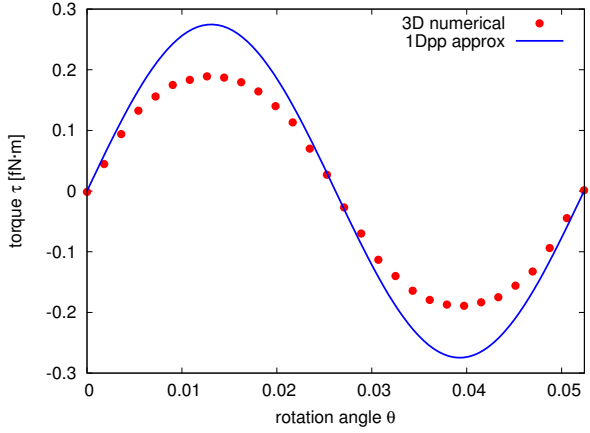}
\caption{Torque as a function of rotation angle for the next-generation \eotwash~apparatus, assuming $\lambda = \beta = 1$ and $\Dzst = 0.1$~mm.  The 1Dpp approximation overestimates the numerical computation of~\cite{Upadhye_Gubser_Khoury_2006} by $\sim 50\%$. \label{f:torque_nextgen_1Dpp}}
\end{center}
\end{figure}

After integrating to find the energy difference $\Delta E$ between the overlapping and non-overlapping configurations, we may proceed as before to find the torque,
\begin{equation}
\tau
=
\Nwedges^2 \fsup \Delta E \sin(\Nwedges \theta).
\end{equation}
This 1Dpp approximation is compared to the three-dimensional numerical calculation in Figure~\ref{f:torque_nextgen_1Dpp}.  1Dpp overestimates the correct torque by $\approx 50\%$.  This is likely due to Approximation~\ref{item:1Dpp_overlapping_region} in Sec.~\ref{subsec:1-D_plane-parallel_approximation}, which artificially flattens out the field in the region $x \gtrsim 0$ in Fig.~\ref{f:source_wedge}, which overlaps the test wedge.  This leads to an overestimate of the energy difference and torque.

\begin{figure}[t]
\begin{center}
\includegraphics[angle=270,width=3.3in]{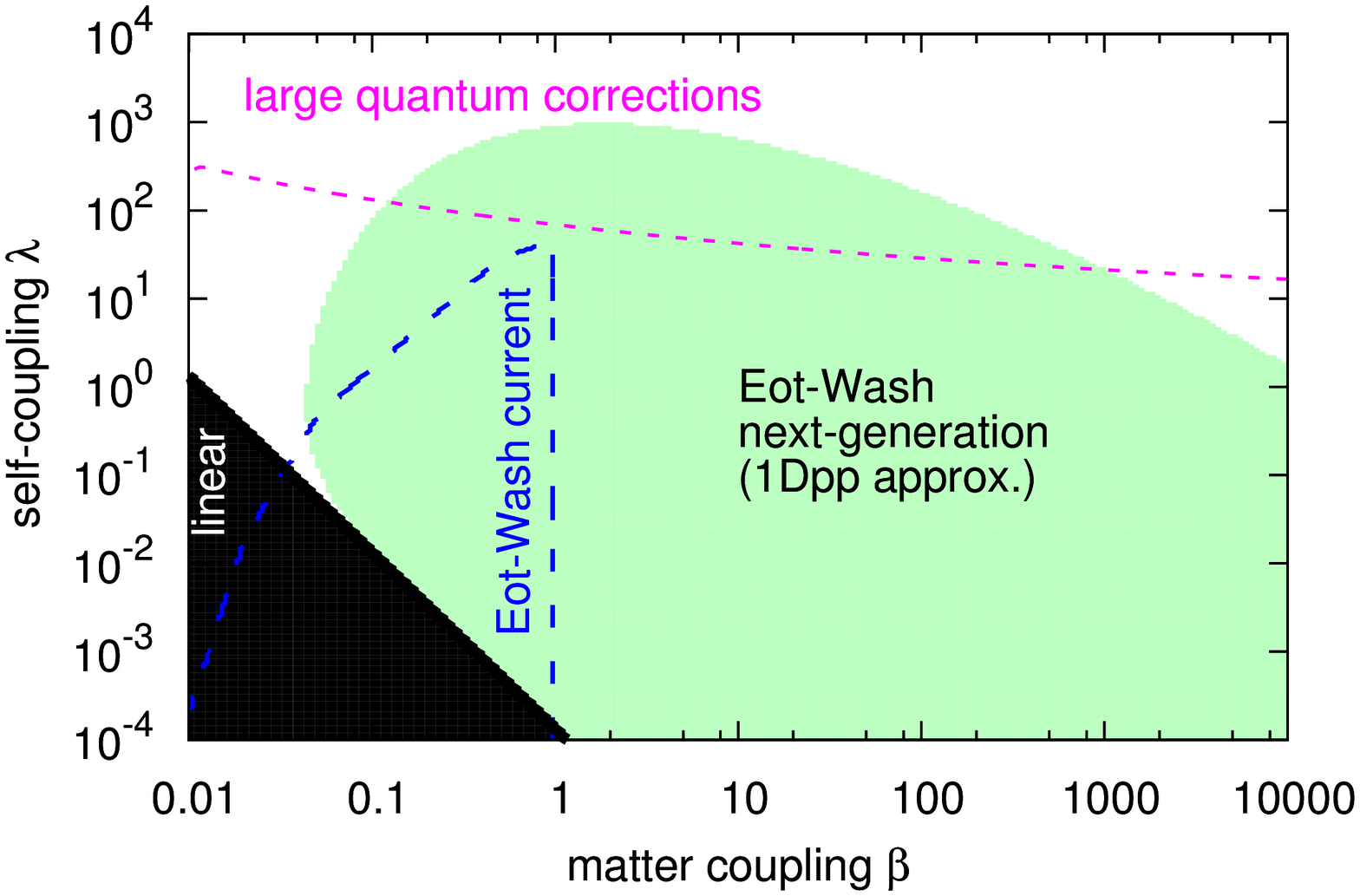}
\includegraphics[angle=270,width=3.3in]{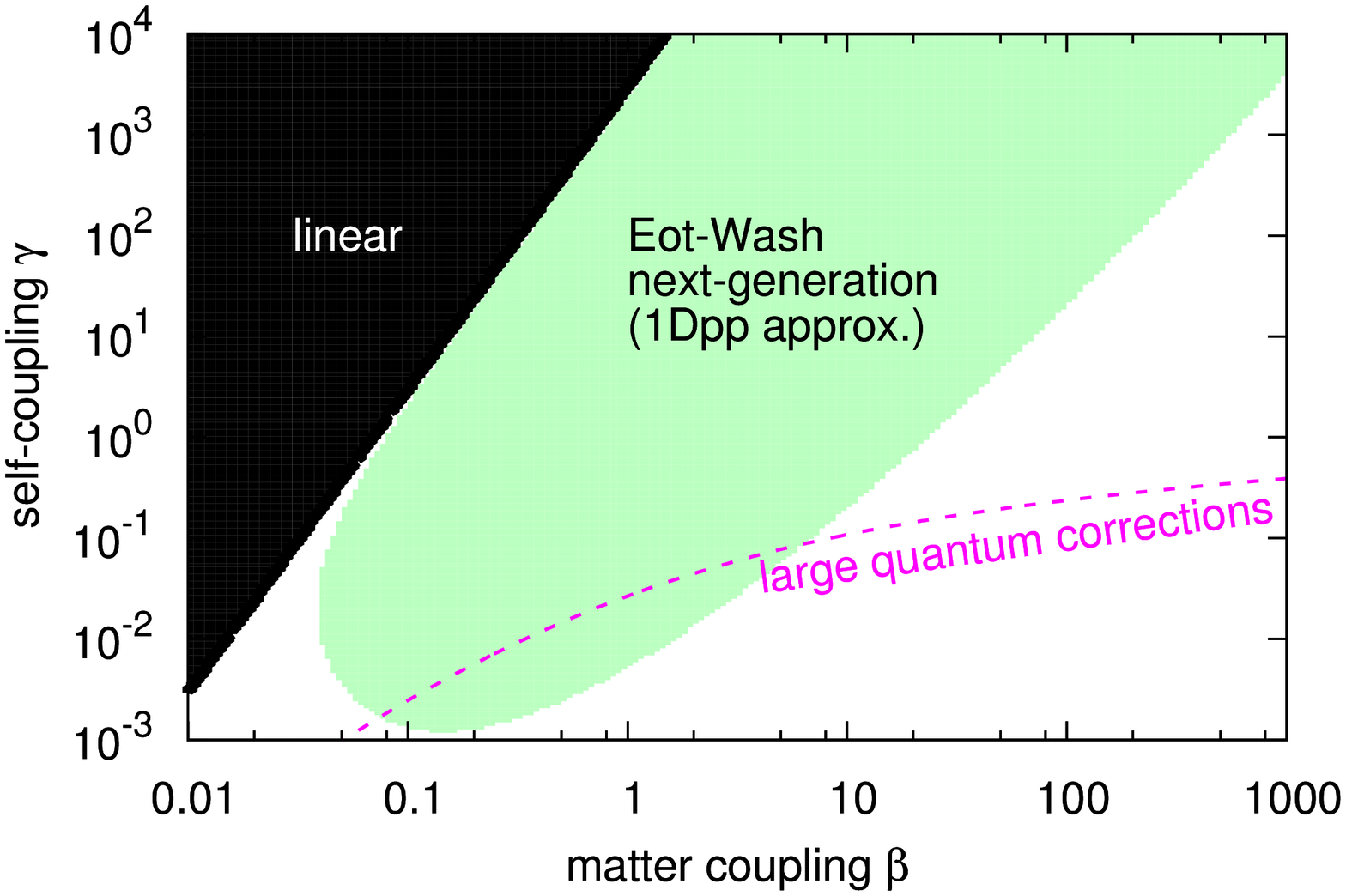}
\caption{Forecast constraints from the next-generation \eotwash~apparatus.  The light green shaded region is excluded; models in the black region are linear inside the disks. 
\Figtop~$n=4$.  Models inside the long-dashed blue curve are excluded by the current \eotwash~experiment, while models above the short-dashed purple line have large quantum corrections.
\Figbot~$n=-1$.  Models below the short-dashed purple curve have large quantum corrections.
\label{f:forecasts_n4_n-1}}
\end{center}
\end{figure}

As with the current experiment, we assume that the next-generation experiment places an upper bound on the torque of $0.003$~fN$\cdot$m at a separation distance $\Dzst = 0.1$mm.  We also assume an identical shielding foil.  The resulting forecasts are shown in Figure~\ref{f:forecasts_n4_n-1}~\Figtop~for the $\phi^4$ potential.  In particular, we note that for a range of matter couplings $0.1 \lesssim \beta \lesssim 1000$, \eotwash~will be able to exclude all $\phi^4$ chameleon models satisfying the quantum stability condition of Sec.~\ref{subsec:quantum_stability_condition}.  This is an improvement of two orders of magnitude relative to the current experiment.

Figure~\ref{f:forecasts_n4_n-1}~\Figbot~forecasts constraints on the $n=-1$ chameleon.  Constraints themselves are not substantially stronger than those of the current experiment.  However, differences in the geometry and the density mean that quantum corrections are larger.  All quantum-stable, nonlinear $n=-1$ chameleons with $0.07 < \beta < 5$ will be excluded by this experiment.

\begin{figure}[t]
\begin{center}
\includegraphics[angle=270,width=3.3in]{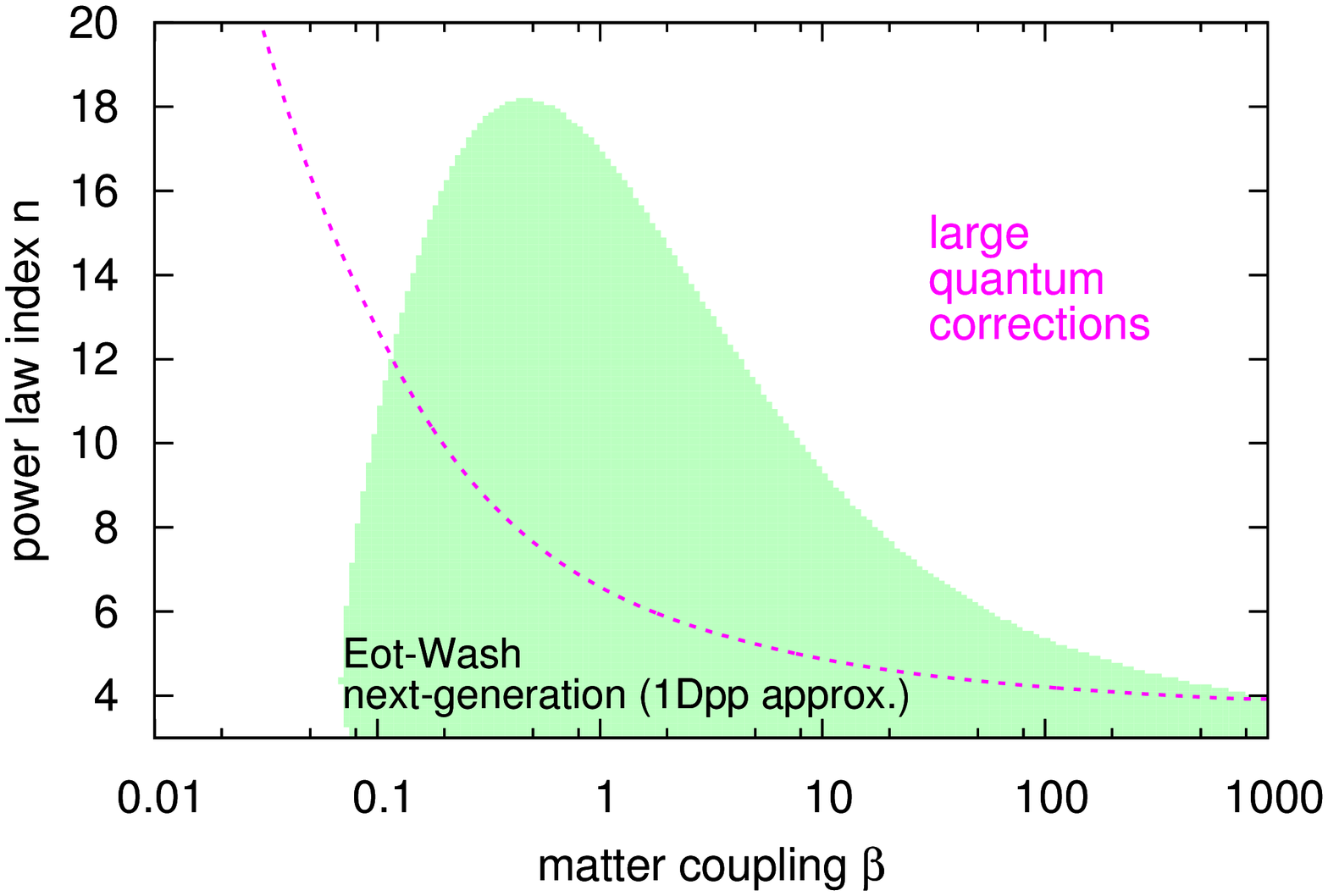}
\includegraphics[angle=270,width=3.3in]{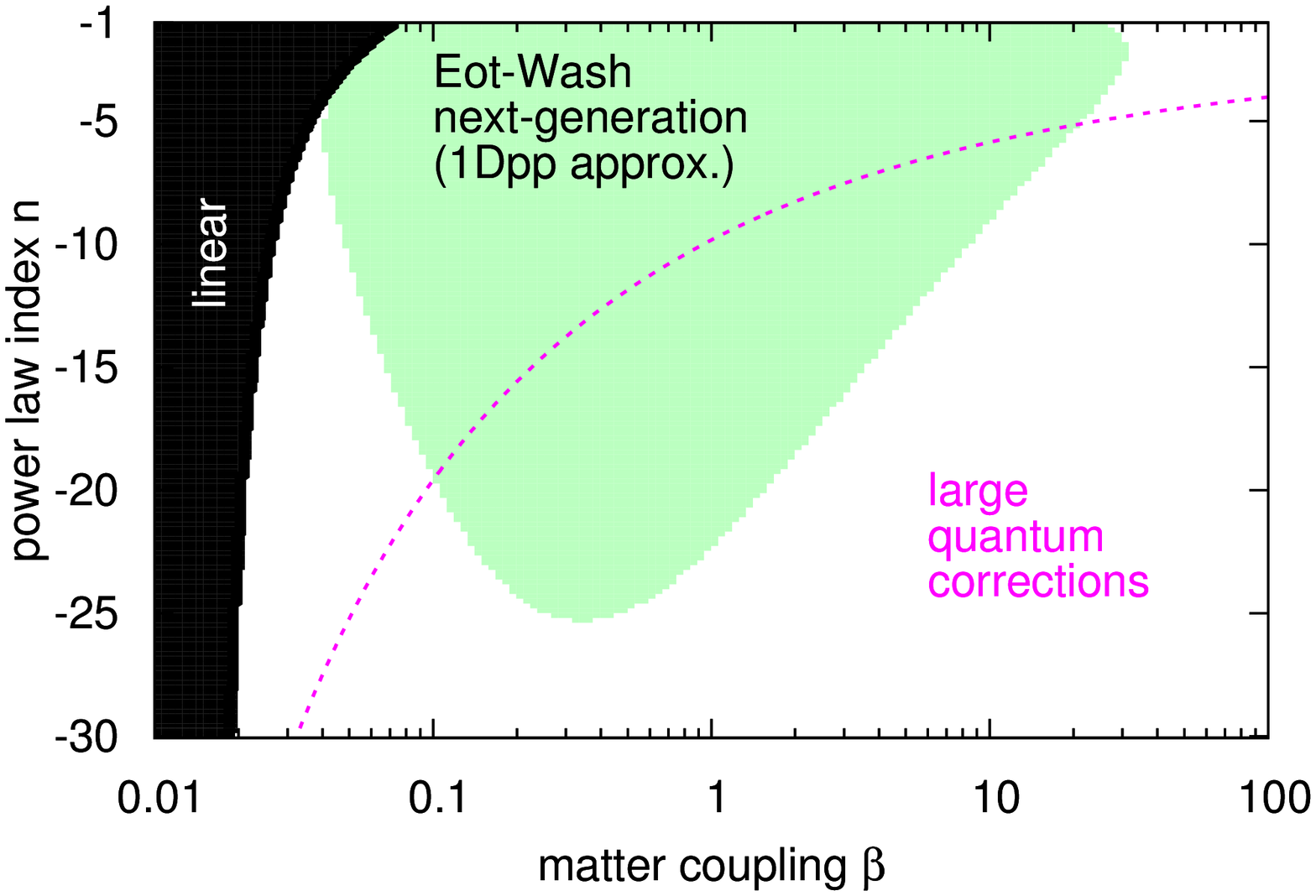}
\caption{Forecast constraints for $\gamma=1$ from the next-generation \eotwash~apparatus.  The light green shaded region is excluded; models in the black region are linear inside the disks. 
\Figtop~Positive $n$.  Models above the short-dashed curve have large quantum corrections.
\Figbot~Negative $n$.  Models below the curve have large quantum corrections.
\label{f:forecasts_gamma_fixed}}
\end{center}
\end{figure}

Constraints on models with large $|n|$ at $\gamma=1$, shown in Figure~\ref{f:forecasts_gamma_fixed}, will improve dramatically relative to those of the current experiment.  In the case of $n>2$, the next-generation \eotwash~will exclude all quantum-stable chameleon models with $0.1 < \beta < 1000$.  For $n \leq -1$, it will exclude all such chameleon models with $0.1 < \beta < 20$.  This is consistent with the claim of~\cite{Upadhye_Hu_Khoury_2012} that an order-unity improvement relative to the current-generation experiment would allow \eotwash~to exclude all quantum-stable chameleon models with matter couplings near unity.

\begin{figure}
\begin{center}
\includegraphics[angle=270,width=3.3in]{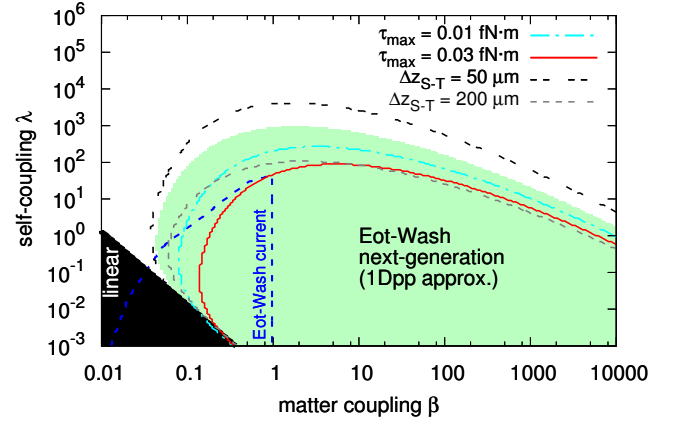}
\caption{Dependence of 1Dpp forecasts on the torque bound (previously assumed to be $\tau_\mathrm{max} = 0.003$~fN$\cdot$m) and the distance probed (previously assumed to be $\Dzst = 100$~$\mu$m).  \label{f:forecasts_range_sensitivity}}
\end{center}
\end{figure}

Since the 1Dpp approximation underestimates the torque in the current experiment but overestimates it in the next-generation experiment, it is possible that the forecasts presented here are an overestimate.  As discussed in the previous section, we could introduce a correction factor to be calibrated using the numerical solutions of~\cite{Upadhye_Gubser_Khoury_2006}.  It is also possible that the experimental sensitivity is somewhat worse than that of the current experiment, or that the next-generation \eotwash~probes a somewhat different distance scale $\Dzst$.  Figure~\ref{f:forecasts_range_sensitivity} shows the constraints which would result for $\phi^4$ theory if the distance or sensitivity differ from our assumed values.  

\section{Conclusion}
\label{sec:conclusion}

Modern torsion pendulum experiments, designed to test Newtonian gravity on submillimeter distance scales, are capable of uncovering new physics at the dark energy scale of $M_\Lambda = 2.4\times 10^{-3}$~eV $\sim (1/0.1\textrm{ mm})$.  We have developed an approximation allowing us to estimate the chameleon-mediated fifth force which would result in a torsion pendulum experiment such as \eotwash~as a function of the chameleon model parameters.  This is accomplished by mapping the geometry of the source and test masses locally onto a one-dimensional plane-parallel problem, which can be solved exactly in a series expansion.  This approximation accurately reproduces the chameleon field on the surface of each mass and allows us to compute the expected torque signal as a function of rotation angle, correct to a factor of $\sim 2$.  Furthermore, it agrees well with published constraints on $\phi^4$ chameleon fifth forces using the current-generation \eotwash~experiment.  

We have used this approximation to extend \eotwash~constraints to a much wider range of chameleon models, as shown in Figures~\ref{f:constraints_n4_n-1} and \ref{f:constraints_gamma_const}.  Of particular interest is the quantum stability condition described in Ref.~\cite{Upadhye_Hu_Khoury_2012} and Sec.~\ref{subsec:quantum_stability_condition}, which argues that current torsion pendulum experiments are on the verge of excluding all chameleon models with small loop corrections and gravitation-strength matter couplings $\beta \sim 1$.  The current experiment can exclude such quantum-stable chameleons for certain specific models, but constraints on them remain weak for inverse power law self-interactions.

Additionally, we have forecast constraints from the next-generation \eotwash~experiment.  This experiment is expected to be powerful enough to exclude a large range of models satisfying the quantum stability condition.
 We show in Fig.~\ref{f:forecasts_n4_n-1}~\Figtop~that the next-generation \eotwash~will exclude all quantum-stable $\phi^4$ chameleons with matter couplings in the range $0.1 \lesssim \beta \lesssim 1000$, an improvement by two orders of magnitude relative to the current experiment.  For unit self-interactions, the next-generation experiment will exclude all quantum-stable $n\geq 2$ models with $0.1 < \beta < 1000$ and all such $n \leq -1$ models with $0.1 < \beta < 20$, as illustrated in Fig.~\ref{f:forecasts_gamma_fixed}.  With the potential to detect or exclude a vast range of quantum-stable power-law chameleon models with gravitation-strength couplings, the next-generation \eotwash~experiment will be a powerful probe of dark energy candidates at the laboratory scale. 

\subsection*{Acknowledgments}

We are grateful to E. Adelberger, T. Cook, F. Fleischer, S. Habib, W. Hu, K. Jones-Smith, J. Khoury, J. Long, J. Steffen, and K. Wagoner for insightful discussions about chameleon theories as well as torsion pendulum experiments. 
The author was supported by the U.S. Department of Energy, Basic Energy Sciences, Office of Science, under contract No. DE-AC02-06CH11357.

The submitted manuscript has been created by
UChicago Argonne, LLC, Operator of Argonne
National Laboratory (``Argonne''). Argonne, a
U.S. Department of Energy Office of Science laboratory,
is operated under Contract No. DE-AC02-
06CH11357. The U.S. Government retains for itself,
and others acting on its behalf, a paid-up
nonexclusive, irrevocable worldwide license in said
article to reproduce, prepare derivative works, distribute
copies to the public, and perform publicly
and display publicly, by or on behalf of the Government.

\bibliographystyle{unsrt}
\bibliography{chameleon}

\end{document}